\documentclass[universe,article,accept,pdftex,moreauthors]{Definitions/mdpi} 

\usepackage{amsmath}

\newcommand{\eb}{\begin{equation}}
\newcommand{\ee}{\end{equation}}

\makeatletter
\let\c@lofdepth\relax
\let\c@lotdepth\relax
\makeatother
\usepackage{subfigure}
\makeatletter
\renewcommand{\@thesubfigure}{\normalsize(\textbf{\alph{subfigure}})}
\makeatother

\usepackage{url}
\definecolor{rkka}{RGB}{219,66,32}
\firstpage{1} 
\pubvolume{1}
\issuenum{1}
\articlenumber{0}
\pubyear{2024}
\copyrightyear{2024}
\externaleditor{Academic Editor: Xian-Yu Wang}
\datereceived{17 August 2024} 
\daterevised{13 September 2024} 
\dateaccepted{16 September 2024} 
\datepublished{ } 
\hreflink{https://doi.org/} 

\Title{Dynamics of Two Planets near a 2:1 Resonance: Case Studies of Known and Synthetic Exosystems on a Grid of \mbox{Initial Configurations}}

\TitleCitation{Dynamics of Two Planets near a 2:1 Resonance: Case Studies of Known and Synthetic Exosystems on a Grid of  Initial Configurations}

\Author{{Valeri Makarov} 
 $^{1,}$*\orcidA{}, Alexey Goldin $^{2}$ and Dimitri Veras $^{3,4,5}$\orcidB{}}

\AuthorNames{Valeri Makarov, Alexey Goldin and Dimitri Veras}

\AuthorCitation{{Makarov, V.;} 
 Goldin, A.; Veras, D.}

\address{%
$^{1}$ \quad {U.S. Naval Observatory}, 
 3450 Massachusetts Ave NW, Washington, DC 20392-5420, USA\\
$^{2}$ \quad {Teza Technology}, 150 N Michigan Ave, Chicago, IL 60601, USA; {alexey.goldin@gmail.com} 
\\
$^{3}$ \quad Centre for Exoplanets and Habitability, University of Warwick, Coventry CV4 7AL, UK; {dimitri.veras@aya.yale.edu}\\
$^{4}$ \quad Centre for Space Domain Awareness, University of Warwick, Coventry CV4 7AL, UK\\
$^{5}$ \quad Department of Physics, University of Warwick, Coventry CV4 7AL, UK}

\corres{Correspondence: {valeri.v.makarov.civ@us.navy.mil} 
}

\abstract{The distribution of period ratios for 580 known two-planet systems is apparently nonuniform, with several sharp peaks and troughs. In particular, the vicinity of the 2:1 commensurability seems to have a deficit of systems. Using Monte Carlo simulations and an empirically inferred population distribution of period ratios, we prove that this apparent dearth of near-resonant systems is not statistically significant. The excess of systems with period ratios in the wider vicinity of the 2:1 resonance is significant, however. Long-term WHFast integrations of a synthetic two-planet system on a grid period ratios from 1.87 through 2.12 reveal that the eccentricity and inclination exchange mechanism between non-resonant planets represents the orbital evolution very well in all cases, except at the exact 2:1 mean motion resonance. This resonance destroys the orderly exchange of eccentricity, while the exchange of inclination still takes place. Additional simulations of the Kepler-113 system on a grid of initial inclinations show that the secular periods of eccentricity and inclination variations are well fitted by a simple hyperbolic cosine function of the initial mutual inclination. We further investigate the six known two-planet systems with period ratios within 2\% of the exact 2:1 resonance (TOI-216, KIC {5437945}, 
 Kepler-384, HD 82943, HD 73526, HD 155358) on a grid of initial inclinations and for two different initial periastron longitudes corresponding to the aligned and anti-aligned states. All these systems are found to be long-term stable except HD 73526, which is likely a false positive. The periodic orbital momentum exchange is still at work in some of these systems, albeit with  much shorter cycling periods of a few years.}

\keyword{{e}xoplanet dynamics; exoplanet systems; orbital evolution}  
\begin{document}

\section{Introduction} \label{int.sec}

Most of the known extrasolar systems with multiple planets are tightly packed and have higher masses than the inner planets of the solar system. This is mostly an observational bias related to the large contribution of the photometric transit method of detection. The secular dynamics of such systems with more powerful and faster interactions between components is a complex and diverse area of research, with significant applications in the theory of extrasolar habitability and planet formation, for example.

The chances of finding a transient, dynamically unstable planetary system are believed to be negligible, unless the system is very young. Therefore, the bulk of the known systems must be long-term dynamically stable. Numerical experiments have revealed that this is not easy to achieve for some stars with Jupiter-mass companions within the estimated parameter space. Furthermore, the pattern of secular evolution of apparently stable systems is different for low-order orbital commensurabilities and general non-resonant configurations~\citep{1999ssd..book.....M}. Given the finite width of each resonant state in the parameter space, the overlap of first- and higher-order resonances leads to a stochastic time evolution of the system \citep{1980AJ.....85.1122W}. Most of the theoretical work in this direction has been undertaken for restricted planar three-body systems, while numerical inference is limited to a number of specific configurations due to the size of the multi-dimensional parameter space sampled. 

In this paper, we investigate several specific systems with two known planets, as well as several synthetic (but representative of the general population) configurations using numerical integration methods, to address the issue of long-term stability, the onset of dynamical chaos, and secular orbital evolution in the vicinity of the main first-order 2:1 mean motion resonance (MMR). Our consideration is not limited to the planar case, with mutual inclinations of the orbits reaching 10\textdegree. We compare the behavior of systems that are sufficiently far from 2:1---but within several percent---with that of six possibly resonant known systems within 2\% of 2:1. 
This sample allows us to investigate the boundary between the regular non-resonant states and the explicit 2:1 resonance. How many known systems are truly in the 2:1 MMR? We begin with the distribution of the nominal period ratios of the two planets $P_2/P_1$ for all 580 known two-planet systems, which have a jagged pattern with prominent peaks and a narrow gap at 2:1. 
Statistical analysis of this sample distribution is placed on solid mathematical footing in Section \ref{sta.sec} with a newly introduced method based on integer fraction rationalization and the hypothesis of a Poisson 
distribution of probabilities for each discrete rationalization. The validity of each period ratio state is quantified in terms of $p$-values. 

In Section \ref{ex.sec}, we investigate the main patterns of non-resonant two-planet systems in the broader vicinity of 2:1 MMR. An orderly, periodic exchange of orbital momentum takes place in such systems. It results in synchronous periodic variation of both eccentricities in the opposite phase, which has been seen from the numerical integration of a number of specific multi-planet systems, e.g.,~\citep{2014ApJ...784..104K, 2014ApJ...780..124M}, as well as in theoretical studies \citep{2013ApJ...778....6Z}. A less-known fact is that mutually inclined orbits are also involved in a similar periodic, anti-phase exchange of inclination with a different period. Both periods of exchange, $P_e$ and $P_i$, are defined by the physical parameters of the two-planet system, including both initial orbital parameters, planet masses, and orbital period ratio. In Section \ref{gap.sec}, we reveal the sensitivity of the eccentricity period $P_e$ to the $P_2/P_1$ ratio on a grid of initial configurations in the range 1.87 through 2.12. We show that the exact $P_2/P_1=2$ commensurability destroys the periodic orbital momentum exchange. In Section \ref{inc.sec}, we demonstrate for the first time the dependence of both osculating elements' periods on the initial mutual inclination of the orbits and find a simple and accurate functional fit for this dependence. A more detailed and object-specific numerical investigation of six known two-planet systems within 2\% of the 2:1 MMR (with their nominal parameters) is presented in Section \ref{2:1.sec}, focusing on the long-term stability, the apparent presence of chaos, the existence of periodic orbital momentum, the periastron and nodal secular motion and libration, the trajectories of the orbital momentum vectors, the range of $P_2/P_1$ variations, and, in some cases, the implications for the observable transit time variation effects. These studies are performed for a grid of initial inclinations. The new findings of this paper and the heterogeneity of near-resonant systems are summarized in Section \ref{con.sec}.

\section{Statistical Analysis of Period Ratios in Two-Planet Systems} \label{sta.sec}

The period ratio distribution of multi-planet systems has attracted considerable attention in the published literature. It probably reflects the planet formation circumstances, as well as the most effective mechanisms of secular dynamical evolution. \citet{2023ASPC..534..863W} summarized the available data for known systems with tightly packed configurations. The sample distribution of adjacent planets' period ratios has a smooth continuum component overlaid with sharp peaks and troughs (their {Figure 5}). 
 The most enigmatic feature is the apparent narrow gap just below 2:1 resonance countered by a narrow peak just above it. This pattern has also been discussed by \citet{2011ApJS..197....8L}, \citet{2014ApJ...790..146F}. On the theoretical side, most newly formed systems emerging from the inward migration in the gaseous disk should be in tight resonant chains, and subsequent encounters spread them out, breaking a significant fraction of the chains \citep{2017MNRAS.470.1750I}. A pile-up of systems near a first-order MMR can be the result of such broken chains, but it is not clear why direct gravitational interaction should furnish the observed asymmetry. Additional mechanisms of resonant chain breaking have been proposed, including turbulence in the disk \citep{2017AJ....153..120B}, intrinsic orbital instability \citep[which 
  works only with more than two tightly packed planets, however]{2012Icar..221..624M}, and tidal dissipation in the host star \citep{2014A&A...570L...7D}. The latter mechanism pushes close systems above the exact MMR commensurabilities, but it requires an insignificant restoring force to have a permanent effect. Our goal in this section is to verify the statistical significance of narrow features in the observed $P_2/P_1$ distribution for all known two-planet systems using a novel rationalization method, which is independent of the commonly used histograms.

\begin{figure}[H]
\includegraphics[width=\linewidth]{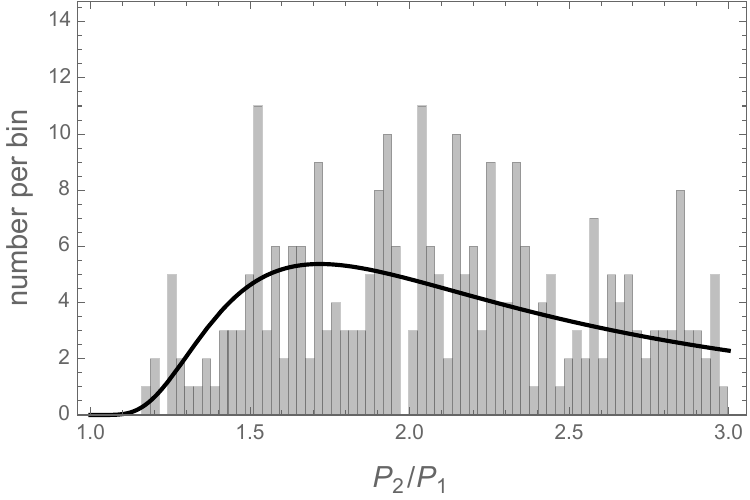}
\caption{{Histogram} 
 of period ratios $P_2/P_1$ for 580 exoplanet systems with two detected planets. The black curve shows
the best fitting analytical distribution of the sample distribution, which is {{\sc FrechetDistribution}}  
$[1.285,1.392,0.842]$.}
\label{pp_hist.fig}
\end{figure}

We begin our analysis with selecting the entire sample of confirmed exoplanets from the NASA Exoplanet Archive available online\endnote{{\url{https://exoplanetarchive.ipac.caltech.edu}}}. 
 Out of 5161 ``published and confirmed'', non-controversial entries in this database with estimated orbital periods (which excludes exoplanets detected by imaging and microlensing) as of August 2023, we select those that belong to systems with two detected planets. The majority of the 3833 systems (2955) are single-planet, while 580 count two planets, 193 count three planets, etc. Orbital period ratios $P_2/P_1$ are computed for the 580 two-planet systems by dividing the longer period by the shorter period. Hence, index 1 refers to the inner planet, and index 2 refers to the outer planet irrespective of their names. The histogram of period ratios in the interval $[1,3]$ in Figure~\ref{pp_hist.fig} shows a notable structure 
 with tightly packed peaks and troughs and a general convex shape. One of the deepest troughs appears to be at 2, corresponding to the first-order mean motion resonance (MMR) 2:1. It is surrounded by prominent peaks on both sides. Taken at face value, we have evidence that exoplanet systems avoid 2:1 MMR and prefer configurations with outer periods slightly longer or slightly shorter than $2\,P_1$.

To confirm or disprove this surmise, we perform a deeper statistical investigation of the observed sample distribution. Given that each bin of $P_2/P_1$ in this histogram counts a relatively small number of systems (maximum 12), we have to test the null hypothesis that the fine structure of the sample distribution is just a random realization of a smooth probability density function (PDF) of a certain population distribution. The regular binning technique with equal widths appears to be an easy and commonly used method, but it produces confusing or discrepant results, with the outcome being sensitive to the technical choice of bin width and positioning. For example, the six exoplanet systems closest to the 2:1 commensurability investigated in Section \ref{2:1.sec} happen to have nominal $P_2/P_1$ values between 2 and 2.017, while no systems are known with $P_2/P_1$ between 1.99 and 2. If we choose a bin width of 0.02, and center the resonant bin on 2, it would include three systems, which is not significantly different from the expected rate. A strong asymmetry (0 against 6) arises only if we place the two symmetric bins of 0.02 around 2. However, as is shown in this paper, at least two partially resonant systems out of the six periodically traverse the 2:1 commensurability because of the finite exchange of orbital energy. Therefore, the apparent asymmetry in the near-resonant systems may be a consequence of the arbitrarily chosen histogram bins rather than an astrophysical fact. To avoid this interpretation ambiguity, we develop here a more sophisticated distribution of discrete rationalizations of $P_2/P_1$. Rationalization is a mathematical function which maps a real number $d$ to a unique ratio of two integers ${\cal R}(d)$ within a specified tolerance interval. The uniqueness of this mapping is achieved by minimizing the sum of the integer nominator and denominator. For example, 2.0 real is obviously rationalized to 2/1, while 2.20 is rationalized to 11/5, although higher-order rationalizations 22/10, 33/15, etc., are also matching. The tolerance input parameter determines how many different rationalizations are obtained for a given data vector. \textcolor{black}{{For example,} 
 2.2111112 is rationalized to 11/5 with a tolerance of 2\%, and the same value is rationalized to to 31/14 with a tolerance of 1\%.}   Using the rationalization intervals instead of the equal-width binning allows us to reproduce a unique and reproducible result.

\begin{figure}[H]
\includegraphics[width=0.48\linewidth]{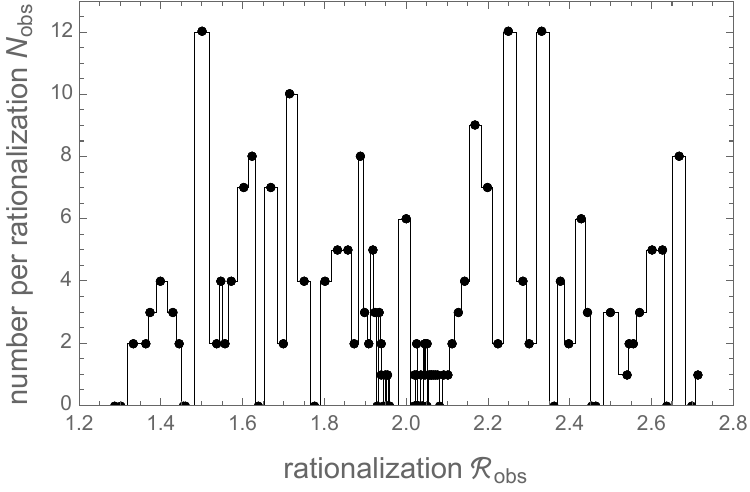}
\includegraphics[width=0.48\linewidth]{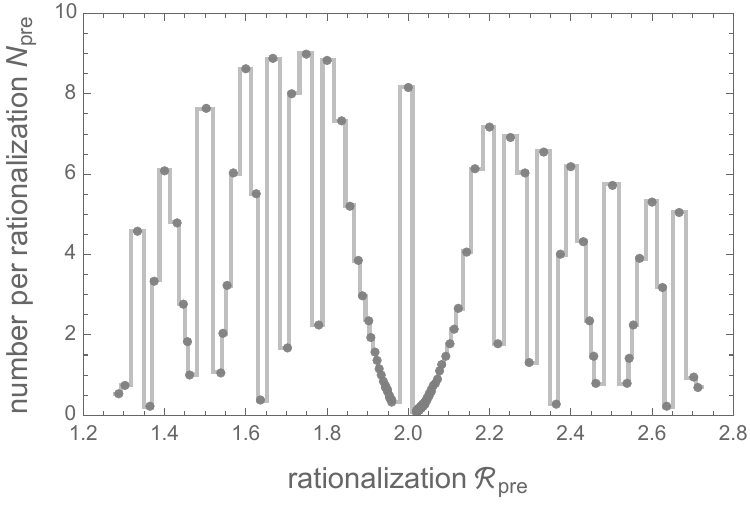}
\caption{Histograms of discrete rationalizations with 2\% tolerance of continuous period ratios $P_2/P_1$. \textbf{Left panel}: observed sample distribution of rationalizations $N_{\rm obs}$ for 251 exoplanet systems with $P_2/P_1$ between 1.3 and 2.7. \textbf{Right panel}: predicted distribution of rationalizations computed from 1.4 million Monte Carlo simulations with an empirical population distribution, normalized to \mbox{251 sample size.}}
\label{rr_hist.fig}
\end{figure}

The novel technique of rationalization binning was applied to 251 known planet systems
with observed period ratios in the interval $[1.3,2.7]$ centered on the 2:1 MMR. A tolerance distance of 0.02 (2\%) was chosen from the consideration of number statistics and significance of the result.
\textls[-15]{This relatively wide tolerance reduces the number of possible rationalizations within the sample range, thus increasing the number of data points per rationalization bin. Since the number of instances per bin in a given sample follows Poisson distribution, we need at least 4 data points to obtain a signal-to-noise ratio of 2. There are 71 distinct rationalizations for the interval $[1.3,2.7]$ with a tolerance distance of 2\%, and statistically meaningful results can be expected for many of them. The rationalized 251 period ratios constitute a quantized data set ${\cal R}_{\rm obs}$. For each possible rationalization (one out of 71), the number $N_{\rm obs}$ of occurrences in the sample is counted. The distribution of these numbers is the analog of the histogram in Figure~\ref{pp_hist.fig} mapped onto  the discrete space of rationalizations. The resulting distribution of rationalizations is shown in \mbox{Figure~\ref{rr_hist.fig}, left~panel.}}

The underlying population distribution of period ratios in binary exoplanet systems is obviously nonuniform, which has to be accounted for in a statistical interpretation. To find the underlying ``smooth'' population probability density function (PDF), we return to the sample distribution shown in Figure~\ref{pp_hist.fig} and select the best-fitting analytical probability distribution with free parameters over the entire data set using the Wolfram Mathematica {\sc FindDistribution} 
 function\endnote{\url{https://reference.wolfram.com/language/ref/FindDistribution.html}}. \textcolor{black}{The optimization fit with free parameters resulted in a Frechet distribution function} {\sc FrechetDistribution}$[1.285,1.392,0.842]$, with the first parameter called shape, the second called scale, and the third called location parameter. The shape above unity signifies that there are no data values close to 1 (i.e., planets cannot be too close). The scale above unity indicates a very slowly declining tail of the distribution. The fitting PDF is shown in Figure~\ref{pp_hist.fig} with the black curve. The maximum probability (mode) is achieved at $P_2/P_1=1.731$. 

A smooth population distribution of $P_2/P_1$ values maps to a strongly structured population distribution over the discrete space of rationalizations. To find this underlying  distribution, a massive Monte Carlo simulation is performed, randomly generating over 1.4 million period ratio values with the empirically determined Frechet distribution. Each of these values is rationalized with 2\% tolerance. The union of this set defines the predicted discrete space of possible rationalizations ${\cal R}_{\rm pre}$, which counts 113 distinct values in the interval $[1.3,2.7]$. We observe that $113-71=42$ predicted elements of ${\cal R}_{\rm pre}$ do not have any matches in the observed set; thus, the corresponding $N_{\rm obs}=0$. The number of instances for each predicted rationalization $N_{\rm pre}$ is counted, and the predicted histogram is re-normalized to the sample size (251). The resulting expected histogram is shown in Figure~\ref{rr_hist.fig}, right panel. We note that certain predicted rationalizations are intrinsically less likely to happen, which is a property of number theory. Most notably, there is a deep trough around 2/1, while the expected value at this rationalization ($N_{\rm pre}(2/1)=8.17$) is close to the upper envelope. Therefore, the observed deficit of instances around 2:1 in the left panel may in fact be consistent with the expectation.

To further quantify the confidence associated with the features in the observed histogram $N_{\rm obs}$, we assume that the rate of instances at a given  rationalization is Poisson-distributed. Then, each $N_{\rm obs}$ is a single-point realization of a {\sc PoissonDistribution}$[N_{\rm pre}]$ process. This allows us to compute the probabilities of deviations $N_{\rm obs}-N_{\rm pre}$ for each specific rationalization. Specifically, the $p$-value of the null hypothesis (that the observed deviation is just a random fluctuation) is computed from the CDF of the Poisson distribution. We are interested to know how likely the observed peaks and troughs of $N_{\rm obs}$ are to occur randomly. For an excess of instances, $N_{\rm obs}>N_{\rm pre}$, the $p$-value is $1-$CDF$(N_{\rm pre})$. For a deficit of instances, 
$N_{\rm obs}<N_{\rm pre}$, the $p$-value is CDF$(N_{\rm pre})$. Conversely, the $1-p$ values quantify the confidence that these features represent statistically significant excess or deficit of planet period ratios. We find that the troughs in the distribution of $N_{\rm obs}$ are not very significant, with the smallest $p$-values 0.054 (6.20 instances expected, 2 observed) at 12/5, 0.055 (9.01 instances expected, 4 observed) at 7/4, and 0.060 (8.85 instances expected, 4 observed) at 9/5. The confidence in other troughs is less than 90\%. Several rationalizations, however, emerge as confident excess from this analysis. These are, in the order of decreasing confidence, 17/9 ($p=0.0035$), 23/12 (0.0058), 7/3 (0.0173), 29/15 (0.0189), 9/4 (0.0255), 3/2 (0.0473), 25/13 (0.0474), etc. The somewhat unexpected conclusion is that period rationalizations in the wider vicinity of 2/1 are more likely to be overabundant than deficient compared to the expectation. This indicates a certain ``pile-up'' of exoplanet period ratios in the vicinity to the 2:1 MMR. The 2/1 rationalization itself, however, is neither statistically abundant nor deficient. With $N_{\rm pre}=8.17$ and $N_{\rm obs}=6$, the $p$-value of deficit is 0.29, which means that the observed number of such systems is close to expectation. In the light of these results, there is no statistical evidence that exoplanet systems avoid the exact 2:1 resonance, but there is statistical evidence that their preferred state is the close proximity to this resonance.

\section{Eccentricity and Inclination Exchange across the 2:1 Resonance}  \label{ex.sec}
Extensive numerical simulations of multiple two-planet systems were performed using the symplectic integrator WHFast \citep{2015MNRAS.452..376R}, which is part of the {REBOUND package} 
 (\url{https://rebound.readthedocs.io/en/latest/}). This code provides high-speed computation in conservative dynamical systems. Most of the computations were made for $3\times 10^5$ yr with 50 steps per period of the inner planet. \textcolor{black}{This duration is much longer than the characteristic times of secular orbital variations in near-resonant systems, allowing us to verify possible drifts and instabilities over a large number of cycles. A few specific systems, e.g., HD 155358, were integrated for 3 Myr to assess the long-term stability and the presence of weak chaos.} The results systems with period ratios outside the 2:1 rationalization revealed that the orbital eccentricity and mutual inclination undergo secular periodic variations with a constant phase shift. If the initial conditions are set to zero eccentricity and inclination of the inner planet (which is called here planet 1) and a finite initial eccentricity and inclination of the outer planet (2), the main mode of secular variations for moderate $e$ and $i$ is very well described by the following {model:} 

\begin{align}
    e_j(t) &= A_e\,|\sin(2\pi t/P_e+\phi_j)|,\label{model1}\\
    i_j(t) &= B_i\,|\sin(2\pi t/P_i+\phi_j)|\label{model1i}
\end{align}

The phase $\phi_1=0$ with these specific initial conditions, and $\phi_2=\pi/2$. This means that when the eccentricity and inclination of one planet is zero, the other planet's eccentricity and inclination are at their
extreme values. It was also empirically determined that the eccentricity exchange periods $P_e$ depend on the planet masses and initial $i_2(0)$ but are weakly dependent on the initial $e_2(0)$ (as long as it is small). The periods $P_e$ and $P_i$ are unequal and non-commensurate, being typically in the range $100{-}1000$ 
 yr for the tested systems. This behavior is consistent with the predictions of the classical Laplace--Lagrange linear perturbation theory, which was developed mostly for the Jupiter--Saturn pair \citep{brouwer1950secular,1999ssd..book.....M}. For nonzero initial parameters of the inner planet, the appropriate functional model becomes
\begin{align}
e_j(t)&=\sqrt{A_1+A_2\,\cos(4\pi t/P_e+2\phi_e+\pi\,j)},\label{model2}\\
i_j(t)&=\sqrt{B_1+B_2\,\cos(4\pi t/P_i+2\phi_i+\pi\,j)}.\label{model2i}
\end{align}

The previously considered special case, $e_1(0)=0$, leads to $A_1=A_2=A_e^2/2$ and  $\phi_e=0$. Equations~\eqref{model2} and \eqref{model2i} then readily transform into Equations~\eqref{model1} and \eqref{model1i}. The model has been proven to work quite well
for tested eccentricities $e_2(0)$ up to 0.1, and initial inclinations $i_2(0)=0,\,1,\,2,\,3,\,5,\,10,\,20$ degrees. 

Integration of specially modified exoplanet pairs with both masses reduced by a factor of 0.1 (but the same other parameters) showed that the periods $P_e$ and $P_i$ are sensitive to the planet's mass. They become nearly 10 times longer. This seems to be consistent with the prediction of the first-order Laplace--Lagrange theory that the eigenfrequency of secular variations is proportional to the square root of the product of planets' masses and inversely proportional to the stellar mass. Earth-like planets in tight systems have longer periods of orbital exchange, but the mechanism is still the same. The question we are addressing now is whether the 2:1 MMR is special with respect to this periodic exchange model.

\section{Exchange of Eccentricity and Inclination versus Period Ratio} \label{gap.sec}

A series of numerical experiments was performed for a synthetic exoplanet system of two planets with the following parameters: $M_s=0.748\,M_{\bigodot}$, $M_1=0.0080\,M_{\rm jup}$, $M_2=0.0111\,M_{\rm jup}$, $P_1=8.181$ d, and the initial parameters at $t=0$: $e_1(0)=0$, $e_2(0)=0.02$, $i_1(0)=0$, $i_2(0)$=5\textdegree,
$\omega_2(0)=0$. Index 1 refers to the inner planet; index 2 refers to the outer planet. \textcolor{black}{The fixed parameters are those of the two-planet system Kepler-165, which represents a typical secularly osculating configuration just outside the 2:1 MMR. The planet masses are in the range of mini-Neptunes, making the secular libration periods longer and allowing us to test the long-term stability of such potentially habitable planets when they cross the resonance. We know from the previous analysis of similar systems that both eccentricities and inclinations undergo mirrored sinusoidal variations with a shift of $\pi/2$.} The outer planet's orbital period is set on a grid of values
$m\,P_1$, $m=1.87, 1.88, \ldots, 2.12$, a total of 26 initial configurations. Each initial configuration was integrated for 300 Kyr with a step of $P_1/50$. The sampled output values of eccentricity, inclination, and other orbital parameters were analyzed for each simulation.

\begin{figure}[H]
\includegraphics[width=0.48\linewidth]{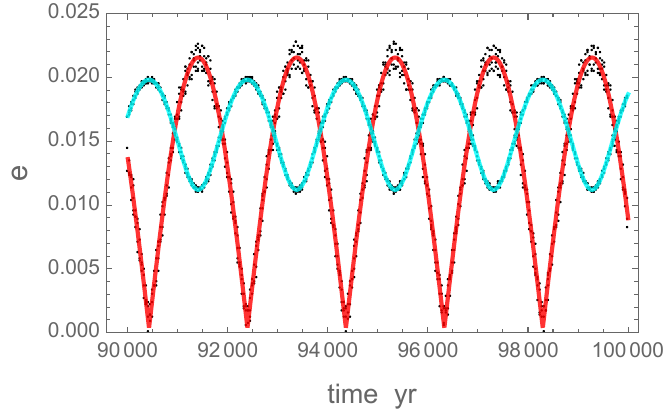}
\includegraphics[width=0.48\linewidth]{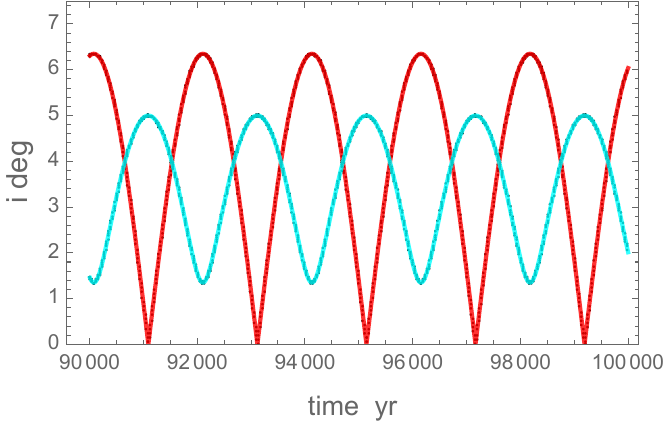}
\caption{{Time} 
 sequences of integrated orbital elements $e$ in the \textbf{left panel} and $i$ in the \textbf{right panel} for a synthetic two-planet system with a period ratio $P_2/P_1=1.98$. The black dots show the actual integration output sampled every 10 yr in the interval between 90 and 100 Kyr. The solid curves represent the optimal fits for these sampled data obtained with Equations~\eqref{model2} and \eqref{model2i}. The red curve in each panel is for planet 1 (inner) and the cyan curve for planet 2 (outer).}
\label{198.fig}
\end{figure}

Figure~\ref{198.fig} shows a typical result of one of such integrations for $P_2/P_1=1.98$, $e_1(0)=0$, $e_2(0)=0.02$, $i_2(0)$ = 5\textdegree, and other parameters as previously described. The left panel shows with black dots the integrated output for $e_1(t)$ and $e_2(t)$. The \textcolor{black}{right panel} shows the simulated results for $i_1(t)$ and $i_2(t)$, also with black dots. A small section of the data spanning 10 yr is displayed. Using nonlinear optimization methods, the best fits were found for  $e_j(t)$ and $i_j(t)$ curves using \textcolor{black}{Equations~(\ref{model2}) and (\ref{model2i})}, which are shown with the red and cyan curves for the inner planet 1 and the outer planet 2, respectively. The cyclic character of variation is quite stable and orderly across the entire span of integration. 
Note the perfectly synchronous modulation of these orbital elements of the two planet with opposite phase. The same type of behavior was found for all these tested configurations except for one special case: the exact 2:1 resonance.

\begin{figure}[H]
\includegraphics[width=\linewidth]{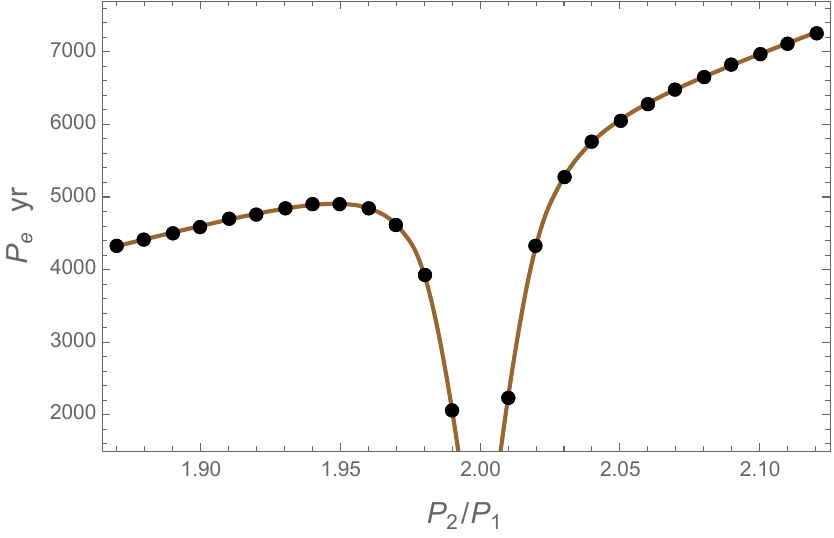}
\caption{Fitted secular periods of eccentricity exchange $P_e$ for \textcolor{black}{Equation~(\ref{model2})} on a grid of orbital period ratios $P_2/P_1=1.87, 1.88, \ldots, 2.12$ obtained from WHFast integrations for 300 Kyr of a synthetic two-planet system.}
\label{gap.fig}
\end{figure}

The estimated $P_e$ values are shown in Figure~\ref{gap.fig}. The dependence of this
parameter on the period ratio is remarkably smooth and generally rising, overlaid with a deep trough at 2:1. The values at $P_2/P_1=1.99$ and 2.01 are nearly half those of the closest neighbors at 1.98 and 2.02. At  $P_2/P_1=2.0$, \textcolor{black}{the model described by Equation~(\ref{model2})} is not applicable because there is no eccentricity exchange. Instead, $e_1$ quickly rises and stabilizes around 0.02, which is the initial value of $e_2$, and continues to vary around that value in a disorderly manner, seemingly forever.

Even though the eccentricity exchange mechanism is broken at the exact 2:1 MMR, the inclinations of both planets still show the same behavior described by \textcolor{black}{Equation~(\ref{model2i})}. We conclude that the 2:1 resonance is not special in terms of the relative orientation of the two orbits, but the main mode of secular eccentricity variation gives way to a set of multiple higher-frequency oscillations, leading to a possibly chaotic trajectory.

\section{Exchange of Eccentricity and Inclination {vs.} 
 Mutual Inclination} \label{inc.sec}
A cautionary note is given in the first-order Lagrange derivation of interplanetary perturbations that it is only applicable to three-body problems with nearly coplanar and nearly circular orbits \citep{1999ssd..book.....M}. Indeed, numerical integrations reveal that the simple model  (Equation~\eqref{model2}) becomes increasingly imprecise  with a wider range of initial eccentricity and inclination. Specifically, we find a strong dependence of both $P_e$ and $P_i$ cycle periods on the mutual inclination of the two planets, which is not predicted by the classical theory. This new property has been confirmed by extensive simulations of a dozen different exoplanet systems, as well as some artificially constructed configurations. 

\begin{figure}[H]
\includegraphics[width=0.48\linewidth]{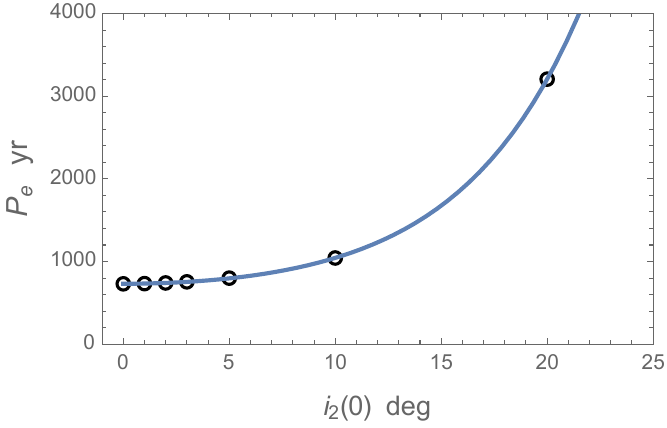}
\includegraphics[width=0.48\linewidth]{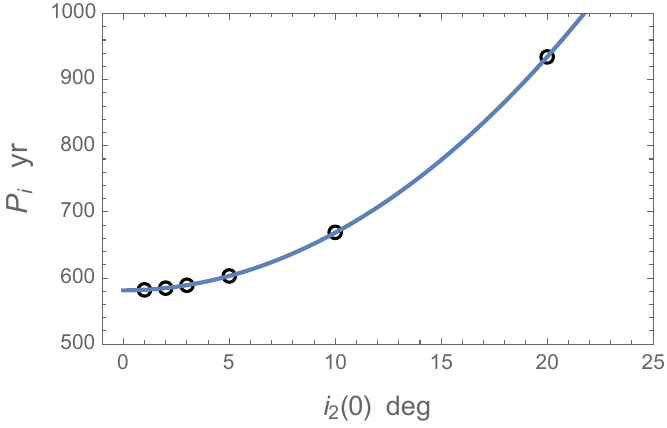}
\caption{Cycle periods of eccentricity (\textbf{left panel}) and inclination (\textbf{right panel}) exchange between the planets b and c in the Kepler-113 system. The results of 14 long-term numerical simulations on a grid of initial inclination are shown with open circles. The curved lines represent the optimal fits obtained with the model $c_0+c_1\,\cosh{(c_2\,i)}$. }
\label{inc.fig}
\end{figure}

Figure~\ref{inc.fig} shows the results obtained for the actual two-planet system Kepler-113. With the periods 4.754 d for planet 1 (b) and 8.925 d for planet 2 (c), the corresponding rationalization of $P_2/P_1$ equals 15/8, which is not remarkable in the statistical sense according to our analysis in Section \ref{sta.sec}. The other assumed parameters are as follows: 
$M_1=3.53\times 10^{-5}\,M_{\bigodot}$, $M_2=2.39\times 10^{-5}\,M_{\bigodot}$, $i_1(0)=0$, $\omega_1(0)=\Omega_1(0)=0$, $\omega_2(0)=\Omega_2(0)=0$, and $e_1(0)=0$. Integration was performed on a grid of initial eccentricity $e_2(0)=0.01,\,0.02,\,0.03,\,0.05,\,0.1$, and a grid of initial mutual inclination $i_2(0)$=0\textdegree,\,1\textdegree,\,2\textdegree,\,3\textdegree,\,5\textdegree,\,10\textdegree,\, 20\textdegree. The eccentricity and inclination exchange mechanism is found for all these test cases. The results displayed in the Figure are for $e_2(0)=0.05$. The output data from each of these 14 simulations for 300 Kyr with a step of 0.095 d is a table of phase space parameters (osculating orbital elements) with a step of 10 yr. We obtain nonlinear optimization fits for the output $e_j(t)$ and $i_j(t)$ using \textcolor{black}{ the model in Equations~\eqref{model2} and \eqref{model2i}}. The sought periods $P_e$ and $P_i$ are the free parameters of this model. The resulting point estimates of the periods are shown with open circles. The solid lines through these points show our empirical fits for $P_e(i)$ and $P_i(i)$, again obtained by nonlinear optimization, using this fitting function:
\eb 
P_e(i)=c_0+c_1\,\cosh{(c_2\,i)},
\label{cosh.eq}
\ee 
and similarly for $P_i(i)$ (with different coefficients). We find that a simple exponential function also provides an adequate---but not perfect---fit.

Additional simulations revealed that the eccentricity exchange mechanism breaks down at initial inclination between 30\textdegree~and 40\textdegree. This can be understood in terms of the empirical model Equation~\eqref{cosh.eq}. The growth of $P_e(i)$ is steep, and the cycle becomes longer than our integration span. The secondary eigenfrequency modes take over and become dominant. The multiple non-commensurate modes combine into a chaotic-looking pattern. However, the $P_i(i)$ exponential rise is much slower. Therefore, even at these high inclination angles, the inclination exchange is evident, and it is well described by Equation~\eqref{model2i}.

\section{A Closer Look at the 2:1 Systems}
\label{2:1.sec}

In the NASA Exoplanet Archive version used for this investigation, six two-planet systems are within 2\% of the exact 2/1 commensurability of periods. These systems are TOI-216, KIC 5437945, Kepler-384, HD 82943, HD 73526, and HD 155358. For all the six systems, the period ratio $P_2/P_1$ is above 2 (but below 2.02). Three systems have been detected by the transit method, and the other three (with HD designations) by the radial velocity method. The physical parameters used in this study for numerical integration of orbits are listed in Table \ref{para.tab}. Eccentricities are not known for two of the transiting systems, so we assumed 0.02 for the inner planet 1 and 0.04 for the outer planet 2 in these cases. The sample includes a variety of planetary masses, eccentricities, and periods, although a dearth of very tight planets is obvious. We performed multiple numerical simulations of the orbital evolution for each of the systems on a two-dimensional grid of initial parameters. The initial argument of periastron $\omega_2(0)$ switched between $0$ and $\pi$ to accommodate the two modes of resonant oscillations previously found for coplanar systems \citep{2008MNRAS.387..747M}, and the initial mutual inclination $i_2(0)$ took values 0\textdegree, 2\textdegree, and 10\textdegree. For each of these six initial configurations, two runs were made for a total duration of 300 Kyr with a dump step of 20 or 200 yr and for 1 Kyr with a dump step of 0.1 yr. The long runs are needed to test for the overall dynamical stability of the systems and to detect possible long-term trends. The short runs were found to be necessary to resolve the actual variability of these near-resonant systems. The main results are summarized below.

\begin{table}[H]
\caption{{Assumed} 
 and adopted physical parameters of six two-planet systems close to 2:1 MMR.\label{para.tab}
}
\begin{tabularx}{\textwidth}{lCCCr}
\toprule
\textbf{Object} & \textbf{Mass} & \textbf{Period, d} & \textbf{Eccentricity}  & \textbf{Provenance} \\
\midrule 
TOI-216 & 0.748 $M_{sun}$ &   &    & db \\
TOI-216.02 & 0.059 $M_{jup}$ & 17.1607  &  0.16 & db \\
TOI-216.01 & 0.56 $M_{jup}$ & 34.5255  &  0.0046  & db \\

\bottomrule
\end{tabularx}
\end{table}

\begin{table}[H]\ContinuedFloat
\small
\caption{{\em Cont.}}
\label{para.tab}
\begin{tabularx}{\textwidth}{lCCCr}
\toprule
\textbf{Object} & \textbf{Mass} & \textbf{Period, d} & \textbf{Eccentricity}  & \textbf{Provenance} \\
\midrule 
KIC 5437945 & 1.07 $M_{sun}$ &   &    & db \\
KIC 5437945 {c} 
 & 0.0821 $M_{jup}$ & 220.13  &  0.02 & * \\
KIC 5437945 b & 0.106 $M_{jup}$ & 440.781  &  0.04 & * \\
\midrule
Kepler-384 & 0.97 $M_{sun}$ &   &    & db \\
Kepler-384 b & 0.00459 $M_{jup}$ & 22.5971  &  0.02 & *  \\
Kepler-384 c & 0.00474 $M_{jup}$ & 45.3483  &  0.04 & * \\
\midrule
HD 82943 & 1.08 $M_{sun}$ &   &    & db \\
HD 82943 c & 1.959 $M_{jup}$ & 220.078  &  0.3663 & db \\
HD 82943 b & 1.681 $M_{jup}$ & 441.47  &  0.162 & db \\
\midrule
HD 73526 & 1.53 $M_{sun}$ &   &    & db \\
HD 73526 b & 3.08 $M_{jup}$ & 188.3  &  0.19 & db \\
HD 73526 c & 2.25 $M_{jup}$ & 379.1  &  0.28 & db \\
\midrule
HD 155358 & 0.793 $M_{sun}$ &   &    & db \\
HD 155358 b & 0.99 $M_{jup}$ & 194.3  &  0.17 & db \\
HD 155358 c & 0.82 $M_{jup}$ & 391.9  &  0.16 & db \\
\bottomrule
\end{tabularx}

\noindent{\footnotesize{Notes: In the Provenance column, db stands for all parameters taken from the Exoplanet Archive, and * stands for assumed eccentricities}}

\end{table}

\subsection{TOI-216 02 and 01}
The two-planet system TOI-216 is remarkable and unique in a few important aspects. Discovered by the transit method from TESS photometric measurements, the system shows an outstanding pattern of transit time variations (TTV) in both planetary companions of the estimated Saturnian mass \citep{2019MNRAS.486.4980K}. Their {Figure 3} 
 shows two statistically significant signals, which emerge as curvature in the residuals of observed transit times with respect to the linear ephemeris model. The TTV signals for the two planets are seen to be in opposite phase, and the time scale of the transit time
variations is of the order of 100 d. A complex dynamical model including 14 parameters was numerically adjusted to these data, which allowed the authors to estimate the orbital eccentricities and masses. The resulting behavior of the orbital elements, however, is not described at all. In the light of the results presented in our Figures~\ref{gap.fig} and \ref{inc.fig}, where the momentum exchange periods for systems farther away from the 2:1 MMR are of the order of 1000 d or longer, we provide a detailed description of the orbital evolution seen from our symplectic integration runs.

\begin{figure}[H]
\includegraphics[width=0.48\linewidth]{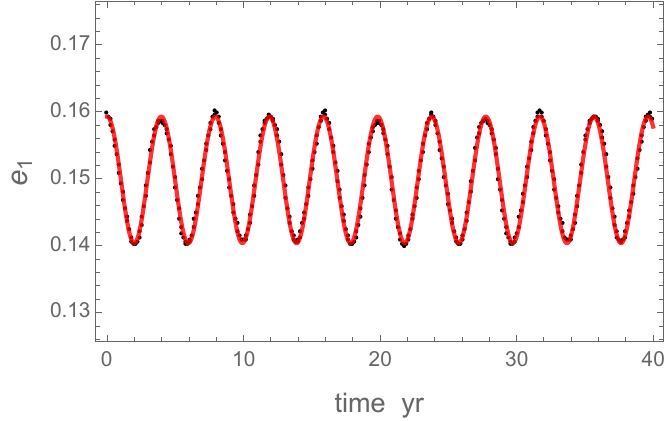}
\includegraphics[width=0.48\linewidth]{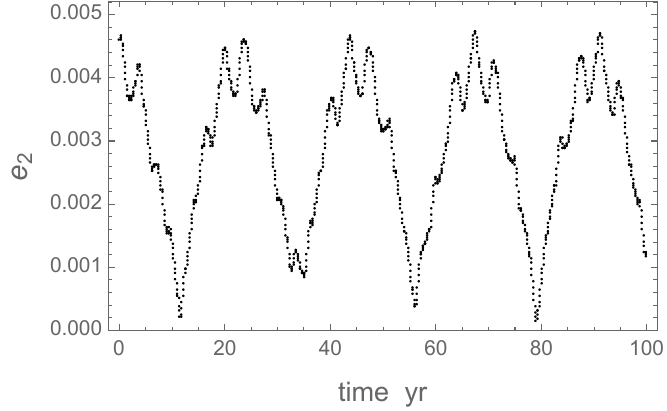}
\caption{Integrated evolution of eccentricity of the inner planet 1 (\textbf{left}) and the outer planet 2 (\textbf{right}) of TOI-216. The red curve for planet 1 shows the best fit obtained by Equation~\eqref{model2}. }
\label{etoi.fig}
\end{figure}

The system TOI-216 is quite stable in these initial configurations across the investigated time span of 300 Kyr. The angular-momentum-related parameters vary in well-defined ranges for both planets without secular trends or catastrophic disruptions. However, we find that the short-term evolution of the osculating elements is significantly different from the previously analyzed systems outside the trough in Figure~\ref{gap.fig}. We will now discuss in more detail the results obtained for one of the tested configurations with the initial conditions $\omega_2(0)=0$, $i_2(0)$=2\textdegree. The other configurations are not too far in terms of the osculating element patterns. The most obvious difference is seen in the evolution pattern of eccentricity (Figure~\ref{etoi.fig}). In comparison with Figure~\ref{198.fig}, left panel, we find that the regular synchronisity of the eccentricity exchange is broken for TOI-216. The more general model in Equation~\eqref{model2} still provides a good fit for the inner planet 1 (shown with red line), but it fails for planet 2. Furthermore, the main frequencies of oscillations are neither equal nor commensurate for the two planets. The estimated $e$-cycle period of the inner planet 1 is 3.963 yr, which is a few orders of magnitude shorter than the exchange periods we have discussed so far. The exact initial period ratio for TOI-216 is $P_2/P_1=2.0119$, which is close to one of the point in the ``gap'' in Figure~\ref{198.fig}. The likely reason for this great reduction of the eccentricity period is the much greater mass of planet 2 than the assumed masses in Section~\ref{gap.sec}. The dominating frequency of $e_2$ oscillations is 22.56 yr, as revealed by a dedicated periodogram analysis of the integration output (not reproduced in this paper for brevity). The dominating frequency of $e_1$ oscillations is seen in the $e_2$ trajectory as a superimposed higher-frequency beat.

The inclination exchange model still works fine for the planets of TOI-216 in the proposed configuration. The common period of inclination exchange for both planets is 335.7 yr. The amplitude of variation is much smaller for planet 2 than for planet 1, which is due to the pronounced asymmetry in planet masses ({Table}
~\ref{para.tab}). Thus, we find a wide dynamical range of characteristic frequencies spanning three orders of magnitude, which describe the orbital variations in this system. How is this related to the exchange of orbital momentum between the planets? To shed light on this mechanism, we computed the orbital momentum vector for each planet at each output state using the instantaneous positions and velocities in the common inertial reference frame. The length of this vector $h$ is a variable of time for each planet because of the momentum exchange. Using a dense grid of 3000 trial periods, we computed amplitude periodograms of $h$ (not shown in this paper for brevity). Despite the variety of characteristic frequencies for $e$ and $i$, both momenta $h_1$ and $h_2$ oscillate with a common period 3.96 yr and with opposite phase. Therefore, \textcolor{black}{ the functional model in Equation~(\ref{model2})} can be successfully applied to $h$ too, with the shorter characteristic period. This fact has interesting implications for the observable transit time variation (TTV) effect.

\begin{figure}[H]
\includegraphics[width=0.88\linewidth]{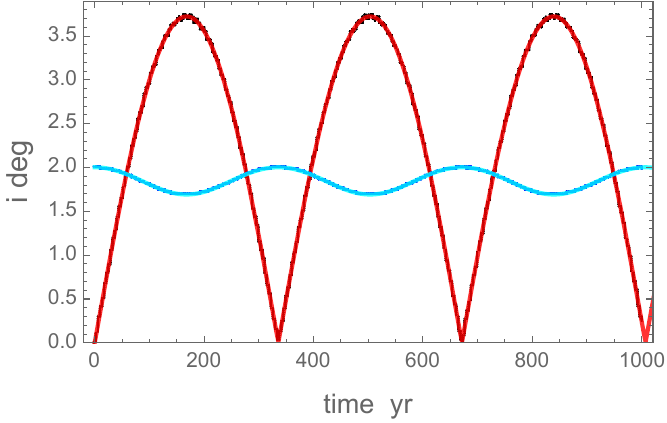}
\caption{{Calculated} 
 and fitted osculating inclinations of the two planets of TOI-216. Integrated states are shown with black dots (barely visible in this graph); the optimal model fits are shown with the red curve for planet 1 (inner) and the cyan curve for planet 2 (outer).  Only a short initial segment of the  WHFast integrations is displayed.}
\label{itoi.fig}
\end{figure}

\begin{figure}[H]
\includegraphics[width=\linewidth]{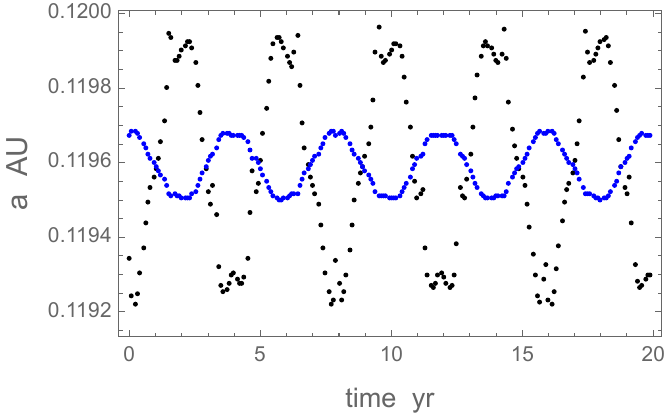}
\caption{Calculated variation of semimajor axis for TOI-216 planets 1 (black dots) and 2 (blue dots) with the initial parameters
$i_2(0)$=2\textdegree, $\omega_2(0)=0$. For the outer planet 2, the values have been reduced by a common offset 0.0706 AU to bring them within the range of the inner planet. Only the starting 20-year-long section of the output is displayed.}
\label{TOI216aa.fig}
\end{figure}

Figure~\ref{TOI216aa.fig} shows the simulated variation of the semimajor axes of TOI-216 planets obtained with the initial configuration $i_2(0)$~=~2\textdegree, $\omega_2(0)=0$. To compare the relative variation, all data points of the outer planet 2 (blue dots) were reduced by 0.0706 AU. The orbits pulsate in unison in the TOI-216 system with the opposite phase, with a much shorter cycle period than the momentum exchange in $i$ (approximately, 4 yr). The amplitudes of this synchronized variation are unequal because of the much greater mass of planet 2. The corresponding full amplitudes of the $P_1$ and $P_2$ oscillation are 0.415\% and 0.065\% relative to the mean periods, or 102.5 min and 32.2 min, respectively. Such variations are indeed confidently detectable from the available TTV measurements. This is consistent with the trajectories presented by \citet{2019MNRAS.486.4980K}, where the system was caught while traversing a maximum in $a_1$ and minimum in $a_2$. 

\begin{figure}[H]
\includegraphics[width=\linewidth]{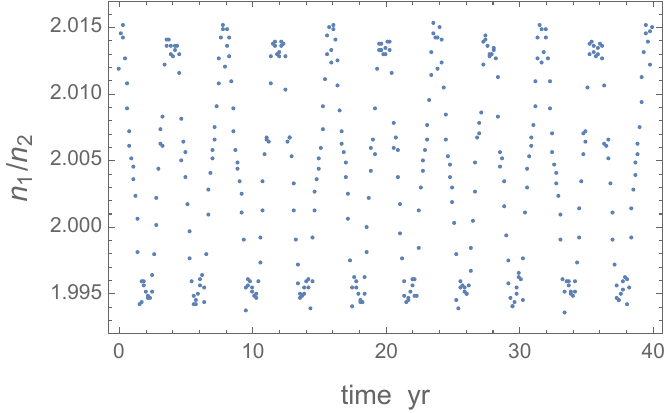}
\caption{Calculated variation of the mean motion ratio $n_1/n_2$ for TOI-216 planets 1 (inner) and 2 (outer) with the initial parameters
$i_2(0)$~=~2\textdegree, $\omega_2(0)~=~0$. Only the starting 40-year-long section of the integration output is displayed.}
\label{TOI216nn.fig}
\end{figure}

The emerging implication of our analysis is that the deviation of the nominal period ratio $P_2/P_1=2.012$ from the exact commensurability has no particular meaning because both periods rapidly and quasi-periodically oscillate within a wider range. 
Indeed, as is shown in Figure~\ref{TOI216nn.fig}, the ratio $n_1/n_2$ of the osculating mean motion rates (and hence the $P_2/P_1$ ratio) rapidly oscillates with a full amplitude of 1\%. The point of exact commensurability is traversed every 2 years. This type of behavior has not been seen for non-resonant systems further away from 2:1 MMR. The apparent narrow gap at 2:1 in period rationalizations, therefore, may be not due to some intrinsic avoidance of the resonance but may simply be a statistical effect caused by the quasi-sinusoidal variation of this ratio around the exact commensurability. The distribution of measured period ratios is bimodal for such systems, with a greater probability to observe the system close to the extrema of the orbital periods than around the time-average period ratio. Continuous TTV observations for a decade should reveal this oscillatory behavior. Note that there are smaller variations on the time scale of months superimposed on the main mode, which may appear as additional noise in TTV measurements on a sparse cadence.

\subsection{HD 155358 b and c}

The two-planet system HD 155358 is an interesting object from the dynamical point of view because divergent results have been presented regarding its chaotic status and long-term stability. The angular momentum deficit  \citep{2017A&A...605A..72L} {(AMD)} 
 method predicts that the system is unstable. The later study by \citet{2023ApJ...943....8G} is based on direct estimation of the minimal Lyapunov time and the related MEGNO parameter within the REBOUND integration package \citep{10.1093/mnras/stw644}. The latter analysis finds a small MEGNO parameter of 2 and a long Lyapunov time of $5.7\times 10^8$ yr. Our simulations of the system generally confirm this conclusion because the system remains intact and stays within a finite range of orbital parameters for all the tested configurations up to 3 Myr. However, our results reveal important detail about this system that have not been previously reported, also explaining the possible cause of the tension.

\begin{figure}[H]
\includegraphics[width=0.48\linewidth]{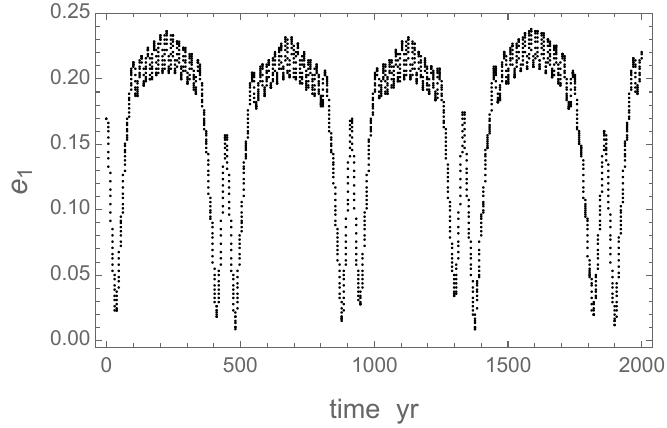}
\includegraphics[width=0.48\linewidth]{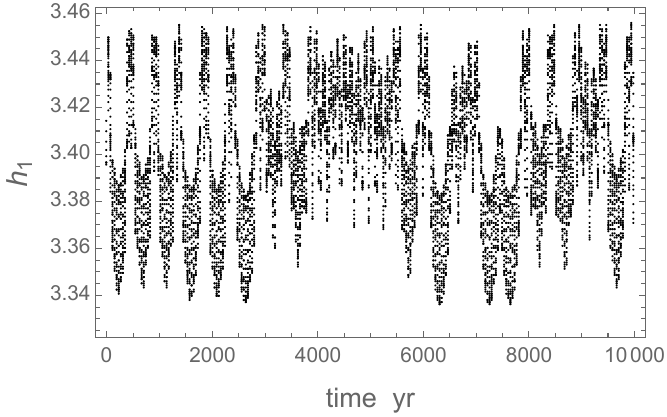}
\caption{{Integrated} 
 evolution of eccentricity (\textbf{left}) and orbital momentum magnitude (\textbf{right}) of the inner planet 1  of HD155358. }
\label{hd15.fig}
\end{figure}

The eccentricity and inclination exchange so explicitly present for TOI-216 is broken for HD 155358. The simulated trajectories for both planets do not follow a simple periodic function and clearly show signatures of dynamical chaos. Figure~\ref{hd15.fig}, left panel, shows the initial segment of the integration of the inner planet with a short dump step (1 yr) and initial $\omega_2(0)=0$, $i_2(0)=0$. Only the first 2000 yr are shown. Rapid variations of $e_1$ suggest that the observed value is transient and should change with time between 0 and 0.24. Although the pattern appears to be regular and quasi-periodic, this is only true for the first 3000 yr. The pattern of variability of all osculating orbital elements dramatically changes after that and becomes rather irregular. This is seen in the right panel, where the magnitude of angular momentum $h_1$ is displayed for a longer stretch. The visual impression is that there are several patterns or orbital trajectories, 
and the system sporadically jumps from one pattern to another on the scale of a few thousand years. We note the relatively large amplitude of $h_1$ variations at $\simeq 1.8\%$. Orbital momentum exchange is definitely on, but it does not have the orderly and periodic character of the non-resonant systems or TOI-216. Rapid variations of orbital elements within 20 yr make the orbital fits from the RV measurements rather uncertain without incorporating osculating components in more complex posterior estimation, as was pointed out by \citet{2001ApJ...551L.109L, 2017MNRAS.469.4613S}.

We find that the anti-aligned coplanar initial configurations with $\omega_2(0)=\pi$ are inherently more orderly and stable than the aligned configurations for this system. A distinct cyclic behavior is seen for both eccentricity trajectories, with $e_1$ varying between 0.07 and 0.20 and $e_2$ between 0.13 and 0.22. On a 1-year cadence, these trajectories display a complex quasi-periodic pattern with multiple wiggles, and the single-mode eccentricity exchange model in inadequate. The periodogram calculation reveals multiple significant modes that combine into this pattern. The greatest modes have periods of 18, 67, 200, and 394 yr. 

Allowing the planets to have mutual orbital inclination adds another level of complexity to this picture. The trajectories in the phase space become obviously chaotic in all the tested configurations, although the system remains intact over 3 Myr. In addition to rapidly varying eccentricities, both inclination angles also vary in a broad range. For example, at $i_2(0)$~=~2\textdegree, both eccentricities vary between 0 and 0.23, while both inclinations vary between 0 and 10\textdegree. The range of inclinations becomes higher at $i_2(0)$~=~10\textdegree~reaching 15\textdegree. However, we detected sporadic periods of a nearly constant inclination at $\sim$5\textdegree~for both planets lasting for approximately 60 Kyr. There is a weakly stable state of this system where the orbits become nearly coplanar, separated by intermittent epochs of chaotic motion of rotation~axes.

\subsection{HD 73526 a and b}
All the tested initial configurations in our numerical simulation resulted in fast disruption of the system. This proposed exoplanet system appears to be inherently unstable. \citet{2014ApJ...780..140W} presented some restricted results with their preferred combination of orbital parameters that were stable for up to $10^8$ yr, but these cases apparently required a lot of fine tuning of the initial relative periastron argument and mean anomaly (their {Figure~6}). 
 \citet{2006ApJ...647..594T} emphasized the importance of dynamical modeling in defining the range of possible orbital solutions for this proposed near-2:1 system. A somewhat stable configuration was also achieved for a coplanar orbital configuration, which is observed edge-on. The latter condition minimizes the estimated planet masses, which is an important factor. At a tilt to the line of sight deviating from 90\textdegree, the increased masses effectively reduce the estimated lifetimes. Given the many limiting conditions required to achieve even a short-term stability, the existence of this two-planet system remains an open issue. We note that the orbital periods are quite close to the previously considered system HD 155358; and yet, the results in terms of stability are drastically different. The implication is that for orbital periods around 200 and 400 d, there is a critical mass of approximately $1\;M_J$ for 2:1 systems to remain stable on time scales comparable to stellar ages.

To alleviate our doubts in the existence of a two-planet system in HD 73526, we performed an enhanced periodogram analysis of the radial velocity data published by \citet{2014ApJ...780..140W}. We produced least-squares periodograms for both AAT/UCLES and Magellan/PFS sets together (fitting an individual zero-point offset) and each set individually. We find that the two instruments produce rather incoherent results in terms of the dominating sinusoidal signals and their relative significance. A peak at 191 d with an amplitude of 70 m s$^{-1}$ is only seen in the combination periodogram, but it is surrounded by two sidelobes of almost equal amplitude. The highest signals from the individual data sets are 362.7 d and 375.0 d, respectively. The results of our limited analysis for the more precise Magellan/PFS collection are adequately consistent with a single sinusoidal signal with a 375.0 d period and 231 m s$^{-1}$ amplitude, although the character of post-fit residuals indicates the presence of a much higher frequency signal or jitter. There is no significant peak around 190 d in the Magellan/PFS data. This is consistent with the theoretical Hamiltonian prediction for HD 73526 in the proposed two-planet configuration, where the outer planet has a high chance of colliding with the inner planet \citep{2017A&A...605A..72L}.

\subsection{HD 82943 b and c}
This system with two planets in the 2:1 period commensurability was determined to be conditionally stable because the chaos indicators are sensitive to initial conditions \citep{2001ApJ...563L..81G}. Long-term stable configurations that are consistent with the observed radial velocity signal are limited in the parameter space to specific multi-dimensional domains. Even for these stable configurations, gravitational interaction between the Jupiter-mass planets causes the osculating elements to rapidly vary on the time scale of a few years, which requires dedicated three-body integration to interpret the observed signal correctly \citep{2004A&A...415..391M}. 

Our numerical results do not agree with the conclusions by \mbox{\citet{2003ApJ...591L..57J}}, who predicted that HD 82943 is locked in either the aligned or anti-aligned MMR. The probable reason for this tension is the different set of planets' physical parameters used in that publication. Specifically, our eccentricity values adopted from \citep{2014MNRAS.439..673B} are significantly smaller. All our long-term simulations spanning 3 Myr with the initial parameters $\Omega_2=0$, $\omega_2=\pi$ produced an immediate disruption of the system precipitated by a boost of one of the eccentricities. This implies that the configurations with initially opposing periastron longitudes are inherently unstable. All the tested initial configurations with aligned periastrons ($\Omega_2=0$, $\omega_2=0$) were found to be long-term stable. This is more consistent with the conclusions by \mbox{\citet{2013ApJ...777..101T}}, who proposed that the system is locked in an aligned 2:1 MMR. However, the emerging orbital evolution in this stable configuration is not consistent with the classical 2:1 MMR, as it is described in \citep{1999ssd..book.....M}.

\begin{figure}[H]
\includegraphics[width=0.45\linewidth]{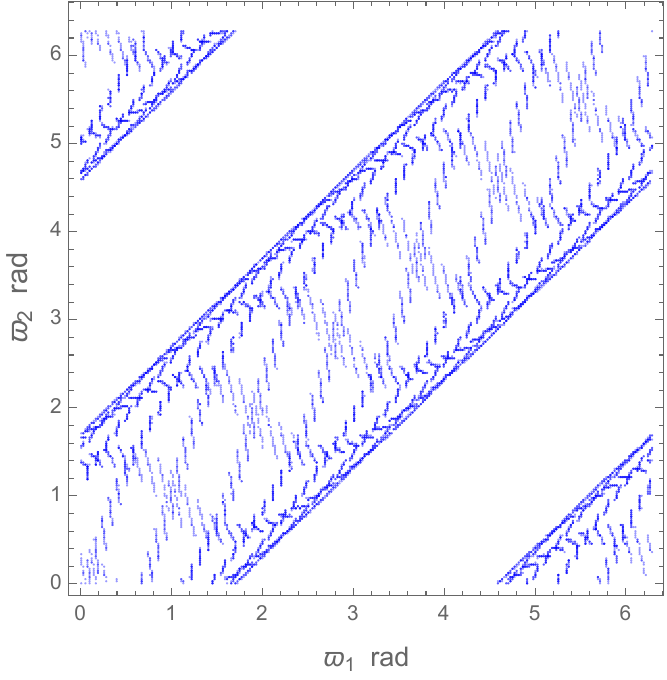}
\includegraphics[width=0.45\linewidth]{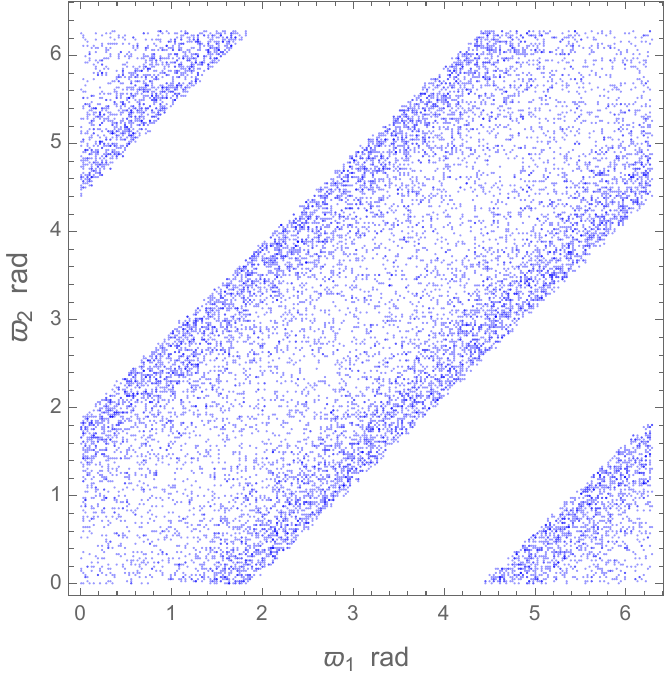}
\caption{Output states of the  HD 82943 system in the $\{\varpi_1,\varpi_2\}$ section. Each dot represents one state sampled every 200 yr. The \textbf{left panel} is for coplanar configuration, $i_2(0)=0$. The \textcolor{black}{\textbf{right panel}} is for a case with significant mutual inclination of orbits, $i_2(0)$~=~10\textdegree.}
\label{pp.fig}
\end{figure}

A deep 2:1 resonance is characterized by the resonant angles
\begin{eqnarray}
    \theta_1&=&\lambda_1-2\lambda_2+\varpi_1\nonumber,\\
    \theta_2&=&\lambda_1-2\lambda_2+\varpi_2\nonumber,\\
    \theta_\varpi&=&\varpi_1-\varpi_2,
\end{eqnarray}
where $\varpi_j=\Omega_j+\omega_j$ is the periapse longitude, and $\lambda_j=\Omega_j+\omega_j+M_j$ is the mean longitude of planet $j=1,2$. These angles should oscillate (librate) around 0 or $\pi$. Resonance is broken if any of these angles circulates, i.e., runs the entire range $[0,2\pi]$ with time. For all tested configurations in this study that are long-term stable, both $\theta_1$ and $\theta_2$ are found to circulate with superimposed oscillations. However, the alignment of periapses holds, as the periastron longitude difference remains within a certain and wide interval centered on 0. A similar case of partial resonance was found by \citet{2020A&A...641A.176P} for the near 3:2 system of K2-19. Figure~\ref{pp.fig} shows the parameter space map in the $\{\varpi_1,\varpi_2\}$ coordinates. The left panel is for the $i_2(0)=0$ (coplanar) configuration, and the right panel is for the $i_2(0)$~=~10\textdegree~case. The overall behavior is similar with the bracketing mutual inclinations in that in both cases, the periastron longitudes remain aligned within $1.7$--$1.8$ rad. Nonzero inclination smothers the pattern of discrete trajectories in the parameter space seen for the coplanar case. Thus, according to this criterion, the system passes the resonance test. However, this figure also shows that both periastron longitudes circulate in unison. Furthermore, when plotted as functions of time, both periastron longitudes show simultaneous  reversals of precession directions separated by $\sim$10$^5$--$10^6$ yr in the coplanar case (not shown in this paper for brevity). Therefore, the lines of apsides remain loosely aligned, as expected in the 2:1 resonance, but they both constantly precess or recess in synchronicity. 

\begin{figure}[H]
\includegraphics[width=0.32\linewidth]{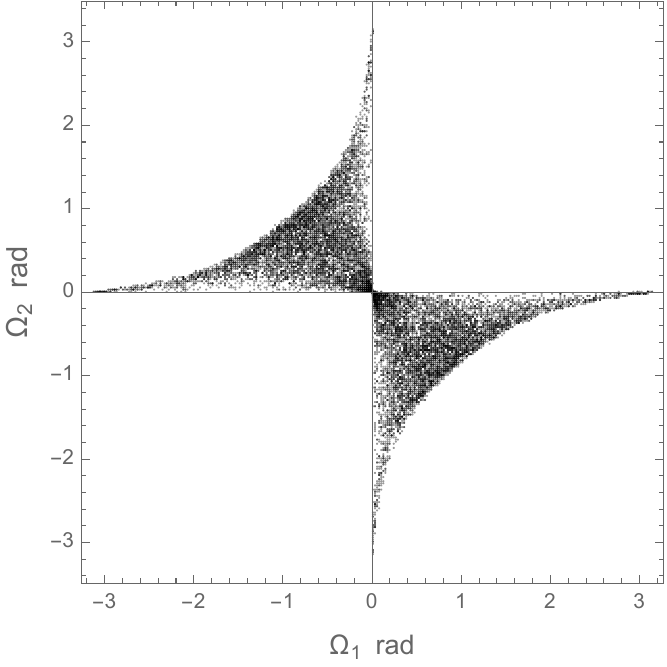}
\includegraphics[width=0.32\linewidth]{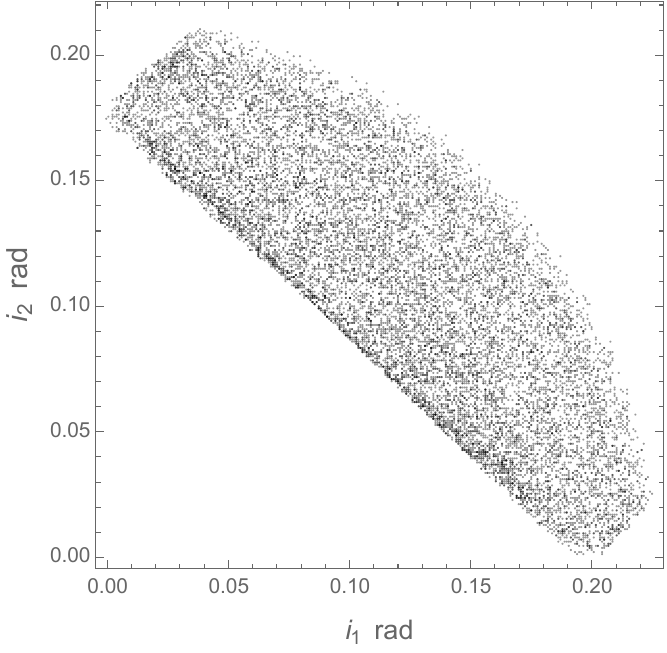}
\includegraphics[width=0.32\linewidth]{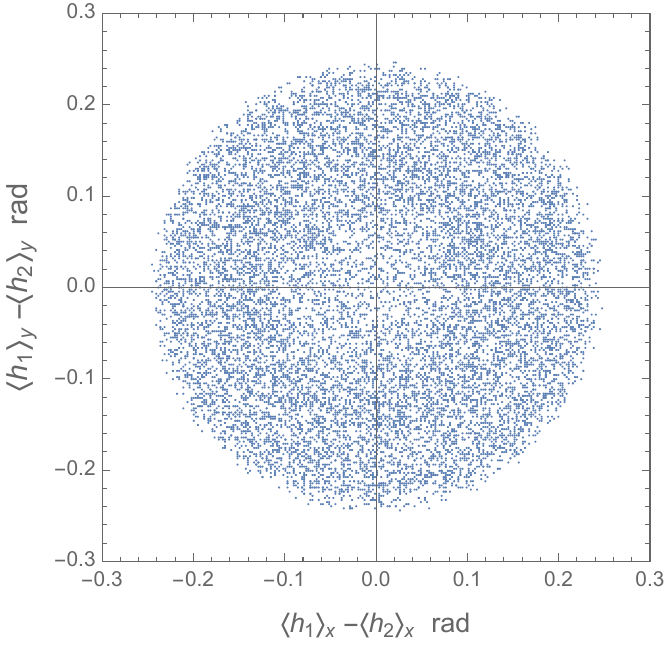}
\caption{Output states of the  HD 82943 system in the $\{\Omega_1,\Omega_2\}$ (\textbf{left panel}), $\{i_1,i_2\}$ (\textbf{middle panel}), and $\{\langle h_1x\rangle-\langle h_2x\rangle,\langle h_1y\rangle-\langle h_2y\rangle\}$ (\textbf{right panel}) sections for the initial configuration with $i_2(0)$~=~10\textdegree. Each dot represents one state sampled every 200 yr. }
\label{hd82.fig}
\end{figure}

Nonzero inclination of orbits brings in an additional level of complexity. The range of varying orbital parameters now includes also $\Omega$ and $i$. The secular behavior of these parameters is correlated too but shows patterns which have not been discussed in the literature to our knowledge. Figure~\ref{hd82.fig} shows the distribution of states sampled every 200 yr with the initial conditions $\omega_2(0)=0$, $i_2(0)$~=~10\textdegree, in various sections. The ascending nodes (left panel) are tangled in a peculiar way so that when one of the nodes circulates through $\pi$, the other is always in the opposite direction at 0. The inclination angles with respect to the $\{x,y\}$ reference plane (middle panel), which is the initial orbital plane of planet 1, display a correlated motion in such a way that both inclinations cannot become closer to 0 than a certain threshold seen as a straight line boundary in the lower left corner. The right panel shows the relative position of the orbit normals projected onto the $\{x,y\}$ reference plane, analogous to the rotational polar movement of Earth. It shows that the angle between the vectors of orbital momentum never exceeds a limit of approximately 0.24 radians. The density of states around $(0,0)$ is lower than the average, signifying that the planets are relatively rarely found with closely aligned orbital momenta.

\begin{figure}[H]
\includegraphics[width=0.45\linewidth]{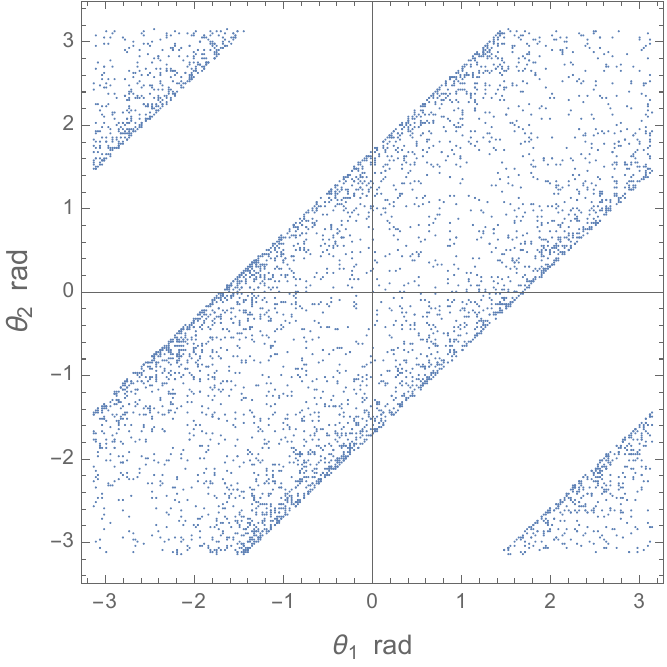}
\includegraphics[width=0.45\linewidth]{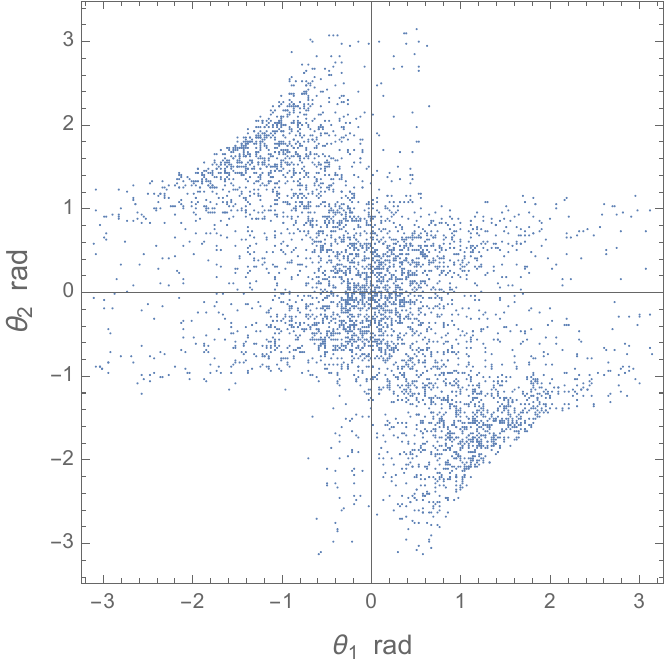}
\caption{Output states of the  HD 82943 system in the $\{\theta_1,\theta_2\}$ section. Each dot represents one state sampled every 200 yr. The \textbf{left panel} is for coplanar configuration, $i_2(0)=0$. The \textcolor{black}{\textbf{right panel}} is for a case with significant mutual inclination of orbits, $i_2(0)$~=~10\textdegree.}
\label{th1th2.fig}
\end{figure}

Further detail on the character of orbital motion in the possible stable configurations of HD 82943 is obtained from the consideration of the two resonance angles $\theta_1$ and $\theta_2$. The classical resonance picture assumes that these two variables oscillate around a fixed value, in this case, 0. Figure~\ref{th1th2.fig} shows that this is not the case. Both angles circulate and cross the entire range. The striking difference between the coplanar case and the inclined case is seen when we compare the the left and right panels. The alignment in planets' positions for the former case (i.e., the libration of $\theta_\varpi$) is broken for nonzero inclination. The empty areas in the corners of the right plot indicate that the planets actually avoid aligned positions at their respective periapses, which are nonetheless loosely aligned. We conclude that the configuration with aligned periastron longitudes represents a new kind of quasi-resonance limited to the relative resonant angle $\theta_\varpi$, which is long-term stable even with finite mutual tilt of the orbits.

\subsection{Kepler-384 b and c}
This system of two transiting planets stands out in our sample of candidate 2:1 resonances by the much smaller masses of the planets. The small perturbing mass effectively reduces the width of the resonance, and the system may be outside the 2:1 MMR with its starting $P_2/P_1=2.0068$. This is what we find from our numerical simulations. The system shows apparently stochastic variation of TTV values with statistically significant deviations up to approximately 200 min \citep{2016ApJS..225....9H}. These variations are not consistent with simple linear or quadratic trends. 

Our long-term integration of the six initial configurations of this system listed in Table \ref{para.tab} (including the assumed small eccentricities) reveal that it is quite stable and orderly and the model of eccentricity/incination exchange applies very well in all these cases, except for the marginal configuration $\omega_2(0)=0$, $i_2(0)$~=~10\textdegree, where additional harmonics of the eigenfrequency with a phase shift become obvious for the inner planet's eccentricity. The cycle periods vary between 8087 and 10,345 yr for eccentricity, and 15,209 and 17,081 yr for inclination, depending on the initial conditions. These long periods of momentum exchange are mostly caused by the small planetary masses. The period ratio $P_2/P_1$ wobbles within a narrow range above the exact 2:1 commensurability, never crossing it. We conclude that this system is outside the true 2:1 MMR, although it shows some characteristics of the previously discussed quasi-resonance.

\begin{figure}[H]
\includegraphics[width=0.32\linewidth]{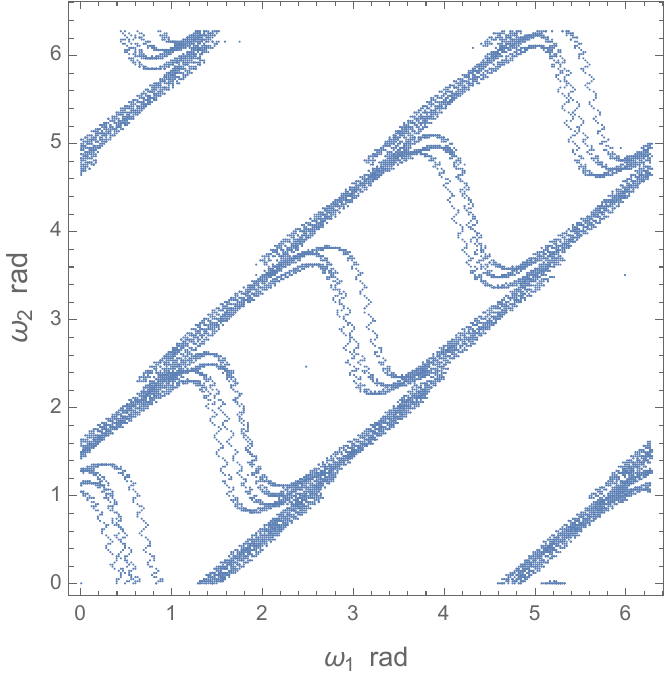}
\includegraphics[width=0.32\linewidth]{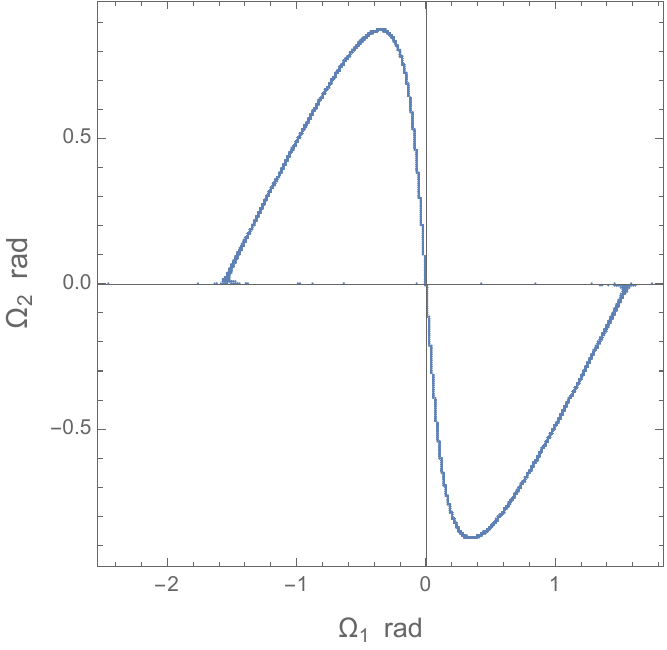}
\includegraphics[width=0.32\linewidth]{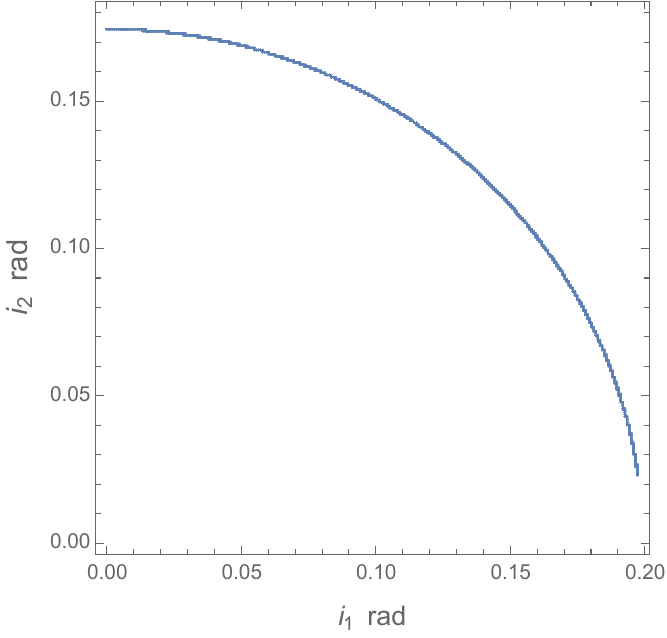}
\caption{Output states of the  Kepler-384 system in the $\{\omega_1,\omega_2\}$ (\textbf{left panel}), $\{\Omega_1,\Omega_2\}$ (\textbf{middle~panel}), and $\{i_1,i_2\}$ (\textbf{right panel}) sections for the initial configuration with $\omega_2(0)=0$, $i_2(0)$~=~10\textdegree. Each dot represents one state sampled every 20 yr. }
\label{kep.fig}
\end{figure}

Figure~\ref{kep.fig} shows the output states of the Kepler-384 system sampled every 20 yr for the marginal initial inclination $i_2(0)$~=~10\textdegree. The periastron arguments $\omega_1$ and $\omega_2$ both circulate along apparently deterministic trajectories (left panel), mostly staying at $\pm$1.4 rad with respect to each other and rapidly switching between these two preferred states. This results in a similar aligned circulation of the resonant angles $\theta_1$ and $\theta_2$. The nodes, on the contrary, move along a very narrow trajectory in such a way that $\Omega_2$ librates around the initial value 0 with an amplitude of 0.9 rad while $\Omega_1$ circulates with a modulated velocity, rapidly passing the phases of aligned ascending nodes and slowing down around the opposite configurations. The inclination angles move strictly along a quarter of a circle (right panel), whose radius is equal to the initial mutual inclination. 

\begin{figure}[H]
\includegraphics[width=0.76\linewidth]{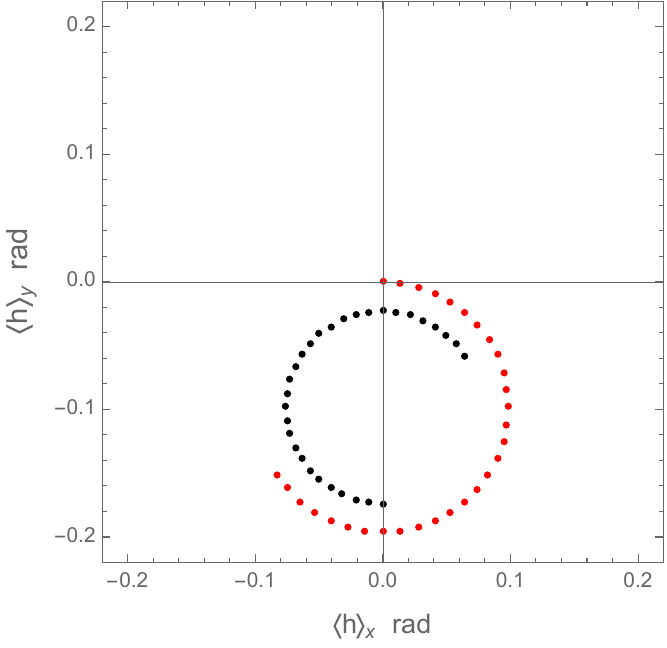}
\caption{Output states of the  Kepler-384 system in the $\{\langle h_x\rangle,\langle h_y\rangle\}$ section, which represent the direction cosines of the orbital momentum vector, for the initial configuration with $\omega_2(0)=0$, $i_2(0)$~=~10\textdegree. Red dots are for the inner planet 1, black dots for the outer planet 2. Each dot represents one state sampled every 400 yr, for the initial 12,000 yr. }
\label{cones.fig}
\end{figure}

As a result of these highly coherent and orderly trajectories, the orbital momentum vectors describe with equal periods two concentric cones in the inertial reference frame, as shown in Figure~\ref{cones.fig}. They move completely synchronously maintaining a constant angle between each other, which is equal to the initial mutual inclination. The axis of the cones is the normal to the invariable plane\endnote{Note that the shift of the circles to negative $\langle h_y\rangle$ is due to the definition of the $(x,y)$ reference plane, which is the initial orbital plane of planet 1.}. This motion is a recession, i.e., a left-hand rotation. Both orbital momentum vectors always remain in a plane rotating around the invariable normal at a constant velocity, also maintaining the initial tilts to this normal. This pattern is quite similar to the second and third Cassini's laws, but apply to the orbital planes of a two-planet system. The consequence of this law is a constant angle between the orbital momentum vectors, which is defined by the initial value. The pattern corresponding to Figure~\ref{hd82.fig}, right panel, is a nearly perfect circle with a radius of 10\textdegree~for this configuration.

Our short-term simulations were used to investigate the high-frequency components of the osculating elements. Following the method of deriving the transit time variations directly from the output states described in \citep{2023Univ....9..463M}, we find that the TTV curves for planet 1 are well approximated by a sinusoid with a period of 18 yr for orbital configurations with initial inclination $i_2(0)$~=~2\textdegree. The higher-frequency variations are nearly negligible. The estimated amplitude is 560 min, so this variation should be detectable with the available measurement precision on the time scale of a decade or longer.

\subsection{KIC 5437945 b and c}
Some historical confusion about this system arises from the dual name in the literature. The initial identification under the KIC 5437945 b name included only one planet b with a period of 440.7813 d \citep{2015ApJ...815..127W}. Follow-up publications listed an additional inner planet Kepler-460 c with a period 220.13 d \citep{2016ApJ...822...86M, 2016ApJS..225....9H}. The planets are listed with different names in the NASA Exoplanet database but with the same host name. We could not find any publications addressing the dynamical status of this system. Our long-term runs for 3 Myr with the 6 initial configurations revealed that the system is stable for this duration, but the aligned periapse configurations (when the two periastron longitudes remain within a certain range of each other) are quasi-periodic and similar to the momentum-exchange model, whereas the anti-aligned configurations are aperiodic and show evidence of secular trends and strong chaos. We conclude that the aligned configuration is the preferred state for this system. The inclination exchange is accurately described by Equation (\ref{model1i}) with a long period of 19 Kyr. On the contrary, the eccentricity exchange takes place on much shorter time scales, and we had to revert to short-time integration with a short dump step to clarify its evolution.

\begin{figure}[H]
\includegraphics[width=0.48\linewidth]{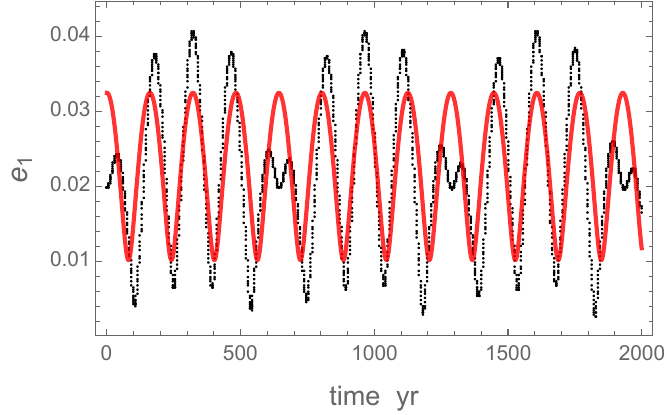}
\includegraphics[width=0.48\linewidth]{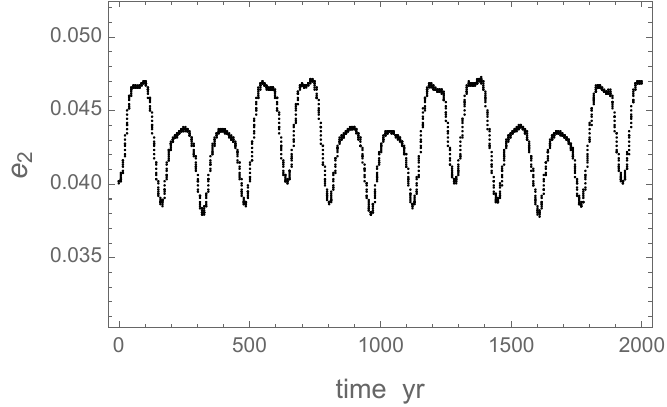}
\caption{Time sequences of integrated eccentricities of the KIC 5437945 system. \textbf{Left panel}: the output osculating eccentricity of the inner planet 1 (planet c) and the best fitting \textcolor{black}{Equation~\eqref{model2}} (red curve). \textbf{Right panel}: the output osculating eccentricity of the outer planet 2 (planet b). The initial configuration $\omega_2(0)=0$ and $i_2(0)=0$.}
\label{k5e1.fig}
\end{figure}

The orbital momentum exchange via eccentricity still takes place for $\omega_2(0)=0$, but it is not represented by a single-mode sinusoidal variation. Figure~\ref{k5e1.fig}, left panel, shows the simulated variation of planet's 1 eccentricity during first 2000 yr for the coplanar case, $i_2(0)=0$. The red curve shows the best fit by Equation~\eqref{model2}, which captures the main period 160.8 yr but fails to reproduce the complex shape of the curve. Planet's 2 eccentricity (right panel) also oscillates in anti-phase with $e_1$ showing wobbles of apparently alternating amplitude. These patterns are similar to light curves of rotating stars with planetary companions observed by {\it {Kepler} 
}, and to obtain a deeper insight into the intricate dynamics of this system, we compute the standard least-squares amplitude periodograms of the osculating orbital elements. Each periodogram was calculated for a grid of trial periods between 60 yr and 800 yr forming a logarithmic cadence of 5000 points. For each trial period, the amplitudes of independent $\sin$- and $\cos$-terms and a separate constant term are computed. 

\begin{figure}[H]
\includegraphics[width=0.32\linewidth]{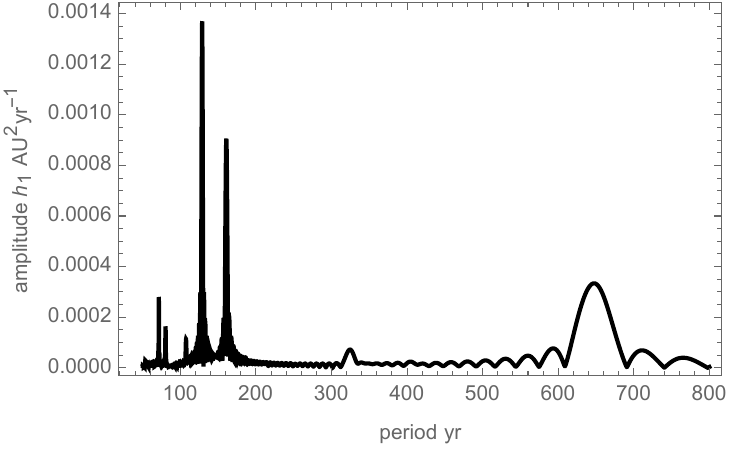}
\includegraphics[width=0.32\linewidth]{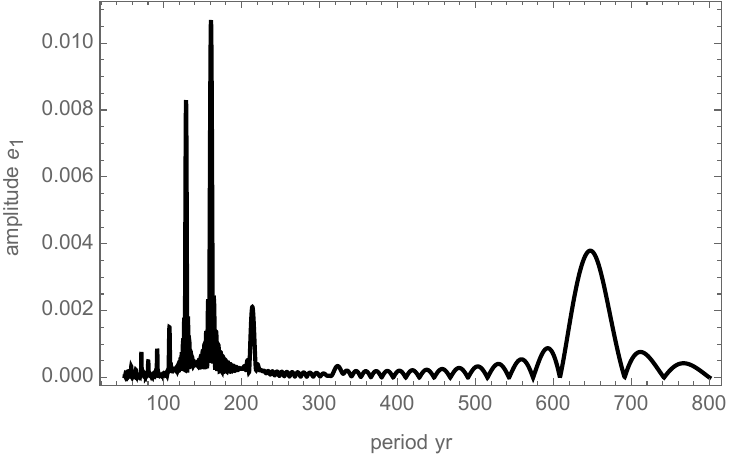}
\includegraphics[width=0.32\linewidth]{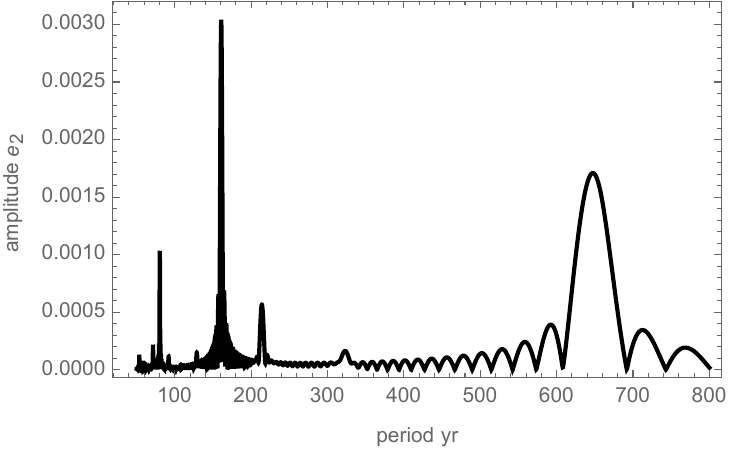}
\caption{Amplitude periodograms of numerically integrated osculating orbital parameters of the two-planet system KIC 5437945 for the
initial configuration $\omega_2(0)=0$, $i_2(0)=0$. }
\label{perio.fig}
\end{figure}

Figure~\ref{perio.fig} displays three of the resulting periodograms: for the inner planet 1 magnitude of orbital momentum $h_1$ (left panel), the inner planet 1 eccentricity $e_1$ (middle panel), and the outer planet 2 eccentricity $e_2$ (right panel). The $h_2$ periodogram is not reproduced because it is practically identical to the $h_1$ periodogram. For planet 2, a couple of modes with periods 128.84 yr and 160.84 yr are dominant. Their interference defines the beating pattern in Figure~\ref{k5e1.fig}. Note that the ratio of these periods rationalizes to 4/5 within 0.2\%, which explains the repeating, quasi-stable features. The relative significance of these two eigenfrequencies is swapped between $h_1$ and $e_1$. In the $e_2$ periodogram, however, the 128.84 yr mode is almost nil, but a fairly significant 80.42 yr mode emerges, which is the exact harmonic of the 160.84 yr mode. In all these periodograms, a prominent modulation with a period of 647.5 yr is present. This may be just the low beating frequency of the two dominant eigenfrequencies. Thus, the main difference of KIC 5437945 from similar systems that are further away from the 2:1 MMR is the interference of two commensurate eigenfrequencies of orbital momentum exchange.

Dynamical chaos in two-planet systems emerges when two or more MMRs overlap~\citep{1980AJ.....85.1122W}. The width of specific resonances in the orbital parameter space greatly increases with eccentricity, as follows from the commonly used model of mathematical pendulum \citep{2018AJ....156...95H}. Therefore, systems with eccentric planets are expected to be mostly chaotic, while systems with small eccentricities, such as the considered configuration of KIC 5437945, may remain non-chaotic even in a close vicinity of a first-order resonance. The pendulum theory, however, is not accurate for the 2:1 MMR with small eccentricities. Our simulations reveal, that this system may formally be outside of the resonance because the output $P_2/P_1$ ratios oscillate in the range 2.002--2.011 (for the aligned periastron longitudes), never actually crossing the resonance point. In terms of the mathematical pendulum model, the system periodically approaches the point of unstable equilibrium hugging the separatrix delineating the zone of stable libration but has not sufficient energy to reach it. The emergence of the secondary prominent eigenfrequency in the spectrum of angular momentum magnitude can be tentatively explained by this proximity to transitory equilibrium state. We also find for this example that the incidence of chaos is dependent, for such transitory systems, on the initial conditions, specifically, the state of periapses' alignment.

\section{Results and Conclusions} \label{con.sec}

The histogram of orbital period ratios for 580 known two-planet systems shows a conspicuous narrow gap just below the exact 2:1 MMR surrounded by prominent sharp peaks on both sides. Is this zone of avoidance real or a mere statistical fluke? We attempted to put the statistical analysis of this sample distribution on a firmer mathematical basis. A smooth empirical population distribution fit in the interval of $P_2/P_1$ values between 1 and 3 yields a Frechet model with modal value around 1.7 and a very slowly declining tail toward higher ratios. This fit is used to quantify the mathematical expectation of the number density, while the observed counts can be assumed the be Poisson distributed around the expected value. To avoid ambiguity related to the width and centering of histogram bins, we computed the confidence intervals of the resulting distribution in fixed rationalizations of the measured ratios within 2\%. Our conclusion is that the narrow gap in the regular histogram is likely to be a coincidence than a significant signal. \textcolor{black}{The apparent bias of resonant systems toward $P_2/P_1>2$ is a mere coincidence too because out of the five such systems (excluding the unreliable HD 73562), at least two, HD 82943 and TOI-216, are found to  traverse the exact commensurability constantly with a period of a decade or shorter.} At the same time, a few peaks in the sample distribution are statistically highly significant, indicating physically preferred commensurabilities at 17/9, 23/12, 7/3, etc. 

To better understand the orbital mechanics of interacting two-planet systems in the broad vicinity of the 2:1 MMR and within this resonance, we performed a number of numerical integrations of synthetic and observed configurations using the symplectic WHFast integrator. We find that the driving mechanism of long-term orbital perturbation is the orbital momentum exchange between the planets, which modulates both eccentricity and inclination in a strictly periodic fashion for non-resonant systems. The phase shift between these oscillations is always $\pi$ such that the minima of one planet's eccentricity or inclination correspond to the maxima of these parameters for the other planet. The temporal behavior is well represented by the first-order Laplace-Lagrange perturbation theory, which was developed to describe long-term perturbations of the solar system's giants. We also revealed previously unknown aspects of this interaction. The periods of momentum exchange are strongly dependent on the initial mutual inclination of the orbits at values above a few degrees. The empirical dependence finds an excellent fit involving a hyperbolic cosine term. The systems with larger amplitudes of relative inclination have significantly longer periods of eccentricity and inclination exchange. An even stronger, non-monotonic dependence on the average $P_2/P_1$ ratio is found (Figure~\ref{gap.fig}) for the eccentricity exchange period. We note that the Laplace--Lagrange perturbation theory generally predicts a monotonically rising $P_e$ with increasing $P_2/P_1$, but it fails to predict the sharp dip in the close vicinity of the 2:1 MMR. At the exact point of resonance, the cyclic momentum exchange mechanism vanishes, being replaced by strong dynamical chaos.

The interplay between the periodic exchange mechanism and the chaotic motion at the exact 2:1 MMR is further investigated via multiple numerical experiments with six listed exoplanet systems within 2\% of this commensurability. The initial configurations spanned a grid of three initial inclination angles (0\textdegree, 2\textdegree, and 10\textdegree) and two opposite alignments of the periapses (0\textdegree~and 180\textdegree). The resulting trajectories for up to 3 Myr were analyzed in terms of (1) overall stability of the system; (2) the presence of a long-period exchange of the orbital momentum magnitude; eccentricity, and inclination; (3) the range of variations of the semimajor axes; (4) the cross-sections of the parameter space in both inclinations, periastron arguments, longitudes of the ascending nodes; (5) the amplitudes and characteristic times of $P_2/P_1$ variations; (6) the spectra of angular momentum magnitudes via periodograms. Within this small sample, we find a variety of complex behavior patterns. All systems are stable or conditionally stable with the exception of HD 73526, which is likely to be a false positive. Our periodogram analysis \textcolor{black}{of this system} shows that there is significant evidence of only one planet in each of the available radial velocity data sets. Conditional stability is mostly defined by the initial configuration of the two periastron directions (aligned and anti-aligned). Strong chaos emerges for some of these systems in one of these configurations, while the orbital variation is more orderly and predictable in the other. The aligned periapses option is more stable for KIC 5437945 and HD 82943, while HD 155358 is less chaotic in the anti-aligned configuration. 

For some systems, such as Kepler-384, regular momentum exchange mechanism is limited to small-inclination conditions. None of the six investigated systems satisfy the theoretical definition of resonance in all the three resonance angles. HD 82943 shows a mixed state, where no libration is found for the angles $\theta_1$ and $\theta_2$, while the third angle $\theta_3$ (relative periapse longitude) does librate within a rather wide range. Kepler-384 shows circulation of both periapses and partial libration of the nodes. In the TOI-216 system, the inner planet shows a clear eccentricity exchange pattern, but the outer planet oscillates in eccentricity out of phase and with a non-commensurate period. These {\it partially} resonant systems
have elevated amplitudes of semimajor axis variations compared to the exchange systems well outside the resonance. Consequently, the period ratio is not constant in time. Kepler-384 and KIC 5437945 remain above the 2:1 ratio but periodically come very close to it. TOI-216, on the other hand, traverses the 2:1 resonance every few years (Figure~\ref{TOI216nn.fig}). Our simulations confirm the TTV in anti-phase for TOI-216 and predict these observable effects for Kepler-384. The near-resonant systems with regular exchange patterns, such as Kepler-384, are found to follow a strict law of orbital plane variation, with the two orbital momentum axes describing concentric cones in the inertial reference frame with the same period. Their orbital axes, therefore, are in one plane with the total orbital momentum vector at any time, maintaining a constant tilt to it.  

\vspace{6pt} 

\funding{{This research received no external funding.} 
}

\dataavailability{{The simulation inputs and results discussed in this paper are available upon reasonable request to the corresponding author.} 
} 

\conflictsofinterest{{The authors declare that the research was conducted in the absence of any commercial or financial relationships that could be construed as a potential conflict of interest.} 
}

\printendnotes

\begin{adjustwidth}{-\extralength}{0cm}

\reftitle{References}

\PublishersNote{}
\end{adjustwidth}
\end{document}